\documentclass[journal]{IEEEtran}

\usepackage[T1]{fontenc}
\usepackage{lmodern}
\usepackage{amsmath,amssymb}
\usepackage{mathrsfs}
\usepackage{bm}
\usepackage{graphicx}
\usepackage[caption=false,font=footnotesize]{subfig}
\usepackage{xcolor}
\usepackage{cite}
\usepackage{algorithm}
\usepackage{algorithmic}
\usepackage{array,booktabs}
\usepackage[colorlinks=true,bookmarksnumbered=true]{hyperref}
\usepackage[capitalise,nameinlink]{cleveref}

\graphicspath{{figures/}}
\definecolor{tacblue}{RGB}{0,0,255}
\hypersetup{linkcolor=tacblue,citecolor=tacblue,urlcolor=tacblue,
  filecolor=tacblue,pdftitle={Three-Dimensional Active Defense of a Surface Target},
  pdfauthor={Taotao Zhao, Zhijian Ji, Lanhao Zhao}}
\newcommand{\doi}[1]{\href{https://doi.org/#1}{doi:\,\nolinkurl{#1}}}

\makeatletter
\def\section{\@startsection{section}{1}{\z@}{3.0ex plus 1.5ex minus 1.5ex}%
{0.7ex plus 1ex minus 0ex}{\normalfont\normalsize\centering\scshape\color{tacblue}}}%
\def\subsection{\@startsection{subsection}{2}{\z@}{3.5ex plus 1.5ex minus 1.5ex}%
{0.7ex plus .5ex minus 0ex}{\normalfont\normalsize\itshape\color{tacblue}}}%
\def\subsubsection{\@startsection{subsubsection}{3}{\parindent}{0ex plus 0.1ex minus 0.1ex}%
{0ex}{\normalfont\normalsize\itshape\color{tacblue}}}%
\makeatother

\makeatletter
\def\@begintheorem#1#2{\@IEEEtmpitemindent\itemindent\relax\topsep 0pt\rmfamily\trivlist%
    \item[]\textit{\indent\textcolor{tacblue}{#1\ #2:}} \itemindent\@IEEEtmpitemindent\relax}
\def\@opargbegintheorem#1#2#3{\@IEEEtmpitemindent\itemindent\relax\topsep 0pt\rmfamily \trivlist%
    \item[]\textit{\indent\textcolor{tacblue}{#1\ #2\ (#3):}} \itemindent\@IEEEtmpitemindent\relax}
\def\@IEEEproof[#1]{\@IEEEQEDshowtrue\par\noindent\hspace{\IEEEproofindentspace}{\itshape\textcolor{tacblue}{#1:} }}
\makeatother

\renewcommand{\eqref}[1]{\textup{\hyperref[#1]{(\ref*{#1})}}}

\newcommand{\tightmath}{\medmuskip=3mu plus 1mu minus 1mu\relax
  \thickmuskip=4mu plus 2mu\relax}
\newcommand{\tightermath}{\medmuskip=2.5mu\relax\thickmuskip=3.5mu\relax}

\newtheorem{theorem}{Theorem}
\newtheorem{lemma}{Lemma}
\newtheorem{corollary}{Corollary}
\newtheorem{proposition}{Proposition}
\newtheorem{definition}{Definition}

\newtheorem{problem}{Problem}
\newtheorem{remark}{Remark}
\newtheorem{example}{Example}

\crefname{equation}{}{}
\crefformat{equation}{#2(#1)#3}
\crefmultiformat{equation}{#2(#1)#3}{ and~#2(#1)#3}{, #2(#1)#3}{, and~#2(#1)#3}
\crefname{figure}{Fig.}{Figs.}
\Crefname{figure}{Fig.}{Figs.}
\crefname{section}{Section}{Sections}
\crefname{subsection}{Section}{Sections}
\crefname{theorem}{Theorem}{Theorems}
\crefname{lemma}{Lemma}{Lemmas}
\crefname{corollary}{Corollary}{Corollaries}
\crefname{proposition}{Proposition}{Propositions}
\crefname{definition}{Definition}{Definitions}
\crefname{assumption}{Assumption}{Assumptions}
\crefname{problem}{Problem}{Problems}
\crefname{remark}{Remark}{Remarks}
\crefname{example}{Example}{Examples}
\crefname{algorithm}{Algorithm}{Algorithms}

\crefrangeformat{equation}{#3(#1)#4\textcolor{tacblue}{--}#5(#2)#6}

\begin{document}

\title{\fontsize{16pt}{19pt}\selectfont\bfseries
  \textcolor{tacblue}{Three-Dimensional Active Defense of a Surface Target}}

\author{Taotao~Zhao,~\IEEEmembership{Student~Member,~IEEE,} Zhijian~Ji,~\IEEEmembership{Senior~Member,~IEEE,}\\
  and~Lanhao~Zhao,~\IEEEmembership{Student~Member,~IEEE}%
\thanks{Manuscript received [date]; revised [date]. This work was supported in part by the National Natural Science Foundation of China under Grant 62373205 and Grant 62033007, in part by the Taishan Scholars Project of Shandong Province of China under Grant tstp20230624, in part by the Taishan Scholars Climbing Program of Shandong Province of China, and in part by the Systems Science Plus Joint Research Program of Qingdao University under Grant XT2024101. \textit{(Corresponding author: Zhijian Ji.)}}%
\thanks{Taotao Zhao and Zhijian Ji are with the Institute of Systems Science, Qingdao University, Qingdao 266071, China (e-mail: 2057798834@qq.com).}%
\thanks{Lanhao Zhao is with Beijing University of Technology, Beijing 100124, China.}}

\markboth{IEEE Transactions on Automatic Control}%
{Three-Dimensional Active Defense of a Surface Target}

\maketitle

\begin{abstract}
In the target--attacker--defender (TAD) game, an attacker pursues a target while a defender tries to intercept the attacker. Existing TAD games are planar, or let all three players move freely in space, where the game still reduces to the plane through the players. This article studies a TAD game in which the target moves on a surface, such as the sea or flat ground, while the attacker and the defender fly above it with non-zero capture radii. Since the target cannot leave the surface, the game no longer reduces to a plane. Firstly, the dominant regions of the target on the surface and of the attacker in space are derived, and the trace of the interception surface on the surface is shown to be an ellipse, a parabola or one branch of a hyperbola. Then, the game of degree is solved: the interception point lies in the air or on the surface, and the target either flees from its foot or runs under it. A strike directly above the target is shown not to be optimal on an open region; there, the target runs under a point in the air, and this crossing in the air also replaces the crossing on the surface wherever the interception surface leans over the path of the target. All solutions are verified through the Hamilton--Jacobi--Isaacs equation. Finally, the barrier is obtained as a level set of the value function. It shows that the collinear construction of earlier work is not semipermeable for a non-zero capture radius, and that a defender nearly above or below the attacker bounds or eliminates the winning region of the attacker. Numerical examples illustrate the results.
\end{abstract}


\section{Introduction}\label{sec:intro}

\IEEEPARstart{S}{hips} and ground vehicles cannot climb away from an aerial threat. Against anti-ship missiles and loitering munitions, they rely on active defense: an interceptor, the defender, is launched to destroy the attacker before it reaches the protected vehicle~\cite{boyell1976defending,karasakal2011antiship}. Such an engagement is a target--attacker--defender (TAD) game. The attacker ($A$) pursues the target ($T$), the defender ($D$) pursues $A$, and $T$ and $D$ cooperate against $A$. In naval and ground defense, the three players do not share the same space: $T$ is bound to the sea or ground surface, whereas $A$ and $D$ fly above it.

Two lines of research address the TAD game. In the first, the players obey linearized kinematics, and the target and the defender design cooperative guidance laws by optimal control or linear--quadratic games~\cite{shaferman2010cooperative,shima2011optimal,perelman2011cooperative,prokopov2013linear,weiss2016minimum,kumar2017cooperative}. In the second, the players obey Isaacs' simple motion~\cite{isaacs1965differential}, and the game of kind and the game of degree are solved in closed form. After the two-team game of Rusnak~\cite{rusnak2005lady}, Pachter et al.~\cite{pachter2014active} and Garcia et al.~\cite{garcia2014cooperative,garcia2015cooperative,garcia2019design} derived state-feedback saddle-point strategies, followed by the complete solution~\cite{pachter2019toward,garcia2021complete}, the fast-defender case~\cite{garcia2017fast} and the superior-attacker case~\cite{garcia2021superior}. Liang et al. constructed the barrier by geometric analysis~\cite{liang2019differential} and studied role switches~\cite{liang2021role} and territorial constraints~\cite{liang2022reconnaissance}. Other studies considered circular targets~\cite{vonmoll2023circular}, unified optimality criteria~\cite{mammadov2024unified}, speed-varying players~\cite{li2026analytical} and anomalous defenders~\cite{wang2026multiagent}, and the geometric solution of a target defense game was justified through the Hamilton--Jacobi--Isaacs (HJI) equation~\cite{fu2023justification}.

These games mostly assume point capture. A capture radius changes the geometry: the Apollonius circle of $T$--$A$~\cite{dorothy2024apollonius} becomes an Apollonius oval~\cite{yan2021lifeline}, and the interception curve of $D$--$A$ recedes by the radius. Wasz et al.~\cite{wasz2019two} solved a two-on-one pursuit with a non-zero capture radius. Shi et al.~\cite{shi2024strategy,shi2025determination,shi2026constructing} then built the barrier of the TAD game with capture radii from the tangency between the Apollonius oval of $T$--$A$ and the limit interception curve of $D$--$A$, along straight strikes in which $A$, its interception point and the capture point are collinear.

All the above games are planar. In three-dimensional space, pursuit--evasion has been studied for aircraft~\cite{miloh1982note,ardema1987three,segal1995barrier,segal1996inertial}, and reach--avoid and capture games have been solved for players that all move in space~\cite{yan2019construction,garcia2020three,yan2022matching,yan2022guarding,chen2024equilibrium,chen2025cooperative}. When all three players of a TAD game move freely in space, the plane through them is a plane of symmetry, and the planar solution applies in this plane. The players of naval and ground defense, however, live in spaces of different dimensions. Li et al.~\cite{li2022different} studied a pursuer in space against an evader on a plane, whose dominant region is cut from an Apollonius sphere by the plane, and Lee et al.~\cite{lee2020perimeter} studied an aerial defender that guards a perimeter against a ground intruder; state constraints of this kind go back to~\cite{plante1971pursuit,pachter1987halfplane}. To the best of the authors' knowledge, no TAD game has been solved for a target on the surface and an attacker and a defender in the air.

Binding $T$ to the surface changes the game in three ways. First, the plane through the players is no longer a plane of symmetry: the optimal interception point leaves it, and the planar solution, lifted to space, is not optimal. Second, the interception surface of $D$--$A$ is a sheet of a hyperboloid, which meets the surface in an ellipse, a parabola or a hyperbola, so that the part of the surface that $A$ can reach may be bounded. Third, $A$ can strike $T$ from directly above, where the payoff is not differentiable and the straight motions toward this point fail. Moreover, for a capture radius $R>0$, the collinear straight strikes of~\cite{shi2025determination,shi2026constructing} are not optimal for $A$, which gains by veering away from $D$ before it strikes, so the barrier built from them is not semipermeable.

This article studies the TAD game in which $T$ moves on a surface, while $A$ and $D$ fly above it with the capture radii $R>0$ and $r>0$. The main contributions are as follows.
\begin{enumerate}
\item Unlike the planar dominant regions of~\cite{garcia2019design,shi2026constructing}, the dominant regions of a surface target and of an airborne attacker are derived. The capture curve of $T$--$A$ is the trace of a Cartesian-oval surface and grows with the height of $A$; the trace of the interception surface of $D$--$A$ is an ellipse, a parabola or one branch of a hyperbola according to the horizontal distance between $A$ and $D$.
\item Compared with~\cite{garcia2019design,shi2026constructing}, where the interception point lies on a planar curve, the game of degree is solved in space. The interception point lies in the air or on the surface, and $T$ either flees from its foot or runs under it. A strike directly above $T$, a degenerate case in the plane, fails on an open region; there, $T$ runs under a point in the air, and this crossing also corrects the crossing on the surface. All solutions are verified through the HJI equation.
\item Unlike~\cite{shi2025determination,shi2026constructing}, the barrier is obtained as a level set of the value function and built as the envelope of two families of disks. The collinear construction is proved not to be semipermeable for $R>0$, and a defender nearly above or below the attacker is shown to bound or to eliminate the winning region of the attacker.
\end{enumerate}

The remainder of this article is organized as follows. \cref{sec:formulation} formulates the game and derives the dominant regions of the players. \cref{sec:degree} solves the game of degree. \cref{sec:kind} obtains the barrier of the game of kind from the value function, and \cref{sec:conclusion} concludes the article.

\section{Problem Formulation}\label{sec:formulation}

\subsection{Game Model}\label{subsec:model}

\begin{figure}[!t]
\centering
\includegraphics[width=\columnwidth]{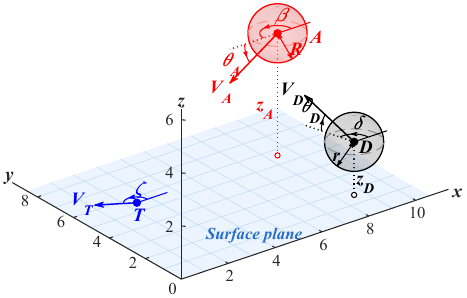}
\caption{TAD game with a surface target. $T$ moves on the surface with heading $\zeta$. $A$ and $D$ fly above it with azimuths $\beta$, $\delta$ and elevations $\theta_A$, $\theta_D$. The balls of radii $R$ and $r$ are the capture regions of $A$ and $D$, and $z_A$, $z_D$ are their heights.}
\label{fig:scenario}
\end{figure}

The Target ($T$), the Attacker ($A$) and the Defender ($D$) obey Isaacs' simple motion~\cite{isaacs1965differential}. Unlike the planar TAD game~\cite{garcia2019design,shi2026constructing}, the engagement takes place in three-dimensional space, as shown in \cref{fig:scenario}. $T$ is a surface vehicle and stays on a plane, such as the sea or flat ground, whereas $A$ and $D$ are airborne. With the $z$-axis pointing upward, the surface and the airspace are
\begin{equation}\label{eq:SH}
\mathcal{S}=\bigl\{(x,y,z)\bigm|z=0\bigr\},\qquad
\mathcal{H}=\bigl\{(x,y,z)\bigm|z\ge 0\bigr\}.
\end{equation}
The positions of the players are $\mathbf{x}_A=(x_A,y_A,z_A)$, $\mathbf{x}_D=(x_D,y_D,z_D)$ and $\mathbf{x}_T=(x_T,y_T,0)$, and $\mathbf{x}=(\mathbf{x}_A,\mathbf{x}_D,\mathbf{x}_T)\in\mathbb{R}^8$ is the state of the game.

The control of $T$ is its heading $\zeta$. The controls of $A$ are its azimuth $\beta$ and elevation $\theta_A$, and those of $D$ are its azimuth $\delta$ and elevation $\theta_D$. Azimuths are measured from the positive $x$-axis to the horizontal projection of the velocity, and elevations from the surface to the velocity; hence $\zeta,\beta,\delta\in[-\pi,\pi]$ and $\theta_A,\theta_D\in[-\pi/2,\pi/2]$. The speeds $V_T$, $V_A$ and $V_D$ are constant. Following~\cite{garcia2019design,shi2026constructing}, $V_A=V_D>V_T$, the speed ratio is $\gamma=V_T/V_A\in(0,1)$, and the speeds are normalized by $V_A=1$. The dynamics $\dot{\mathbf{x}}=\mathbf{f}(\mathbf{x},\beta,\theta_A,\delta,\theta_D,\zeta)$ read
\begin{equation}\label{eq:dynamics}
\begin{alignedat}{3}
\dot{x}_A&=\cos\theta_A\cos\beta, &\;\; \dot{y}_A&=\cos\theta_A\sin\beta, &\;\; \dot{z}_A&=\sin\theta_A\\
\dot{x}_D&=\cos\theta_D\cos\delta, &\;\; \dot{y}_D&=\cos\theta_D\sin\delta, &\;\; \dot{z}_D&=\sin\theta_D\\
\dot{x}_T&=\gamma\cos\zeta, &\;\; \dot{y}_T&=\gamma\sin\zeta, &\;\; \dot{z}_T&=0
\end{alignedat}
\end{equation}
with $\mathbf{x}_A(t_0)=\mathbf{x}_{A_0}$, $\mathbf{x}_D(t_0)=\mathbf{x}_{D_0}$ and $\mathbf{x}_T(t_0)=\mathbf{x}_{T_0}$, where $t_0$ is the initial time. Neither $A$ nor $D$ can pass below the surface, so that
\begin{equation}\label{eq:constraint}
z_A(t)\ge 0,\qquad z_D(t)\ge 0,\qquad t\ge t_0,
\end{equation}
that is, $\theta_A\ge 0$ whenever $z_A=0$, and $\theta_D\ge 0$ whenever $z_D=0$. This state constraint is of the kind met in pursuit--evasion in a half plane~\cite{pachter1987halfplane}.

$A$ pursues $T$, and $D$ protects $T$ by intercepting $A$. $A$ captures $T$ within the radius $R>0$, and $D$ intercepts $A$ within the radius $r>0$; both capture regions are balls. Let $\|\cdot\|_2$ denote the Euclidean norm. The game terminates on $\mathcal{M}=\mathcal{M}_1\cup\mathcal{M}_2$, where
\begin{equation}\label{eq:M}
\begin{aligned}
\mathcal{M}_1&=\bigl\{\mathbf{x}\bigm|\|\mathbf{x}_A-\mathbf{x}_T\|_2\le R\bigr\}\\
\mathcal{M}_2&=\bigl\{\mathbf{x}\bigm|\|\mathbf{x}_A-\mathbf{x}_D\|_2\le r\bigr\}
\end{aligned}
\end{equation}
represent the capture of $T$ and the interception of $A$, respectively. The terminal time $t_f$ is the first time at which $\mathbf{x}\in\mathcal{M}$, and $\mathbf{x}_f=(\mathbf{x}_{A_f},\mathbf{x}_{D_f},\mathbf{x}_{T_f})$ is the terminal state. All players know \crefrange{eq:SH}{eq:M}, the speeds and the current state, so the game has perfect information.

As in combat games with two terminal sets~\cite{ardema1985combat}, each initial state leads to one of three outcomes: $\mathbf{x}$ enters $\mathcal{M}_1$ first, and $A$ wins; $\mathbf{x}$ enters $\mathcal{M}_2$ first, and the $D$--$T$ team wins; or $\mathbf{x}$ enters both at the same time. The states from which $A$ can enforce the first outcome, whatever the $D$--$T$ team does, form the winning region $\mathcal{R}_A$ of $A$. The states from which the $D$--$T$ team can keep $\mathbf{x}$ out of $\mathcal{M}_1$ until it enters $\mathcal{M}_2$, whatever $A$ does, form the winning region $\mathcal{R}_D$ of the $D$--$T$ team. The barrier $\mathcal{B}$ separates $\mathcal{R}_A$ from $\mathcal{R}_D$.

\begin{problem}[Game of Kind]\label{prob:kind}
For an initial state $\mathbf{x}_0$ satisfying \eqref{eq:constraint}, determine whether $\mathbf{x}_0\in\mathcal{R}_A$ or $\mathbf{x}_0\in\mathcal{R}_D$, and construct the barrier $\mathcal{B}$.
\end{problem}

\begin{problem}[Game of Degree]\label{prob:degree}
For $\mathbf{x}_0\in\mathcal{R}_D$, find the saddle-point strategies $(\beta^*,\theta_A^*)$ and $(\delta^*,\theta_D^*,\zeta^*)$ of
\begin{equation}\label{eq:value}
V(\mathbf{x}_0)=\min_{\beta,\,\theta_A}\;\max_{\delta,\,\theta_D,\,\zeta}\;\|\mathbf{x}_{A_f}-\mathbf{x}_{T_f}\|_2
\end{equation}
subject to \eqref{eq:dynamics} and \eqref{eq:constraint}.
\end{problem}

In \cref{prob:degree}, $A$ aims to be intercepted as close to $T$ as possible, while the $D$--$T$ team seeks to keep $A$ away from $T$. Since $T$ stays on $\mathcal{S}$, the squared payoff is
\begin{equation}\label{eq:payoff}
\|\mathbf{x}_{A_f}-\mathbf{x}_{T_f}\|_2^2=(x_{A_f}-x_{T_f})^2+(y_{A_f}-y_{T_f})^2+z_{A_f}^2.
\end{equation}
Compared with the planar payoff of~\cite{shi2026constructing}, \eqref{eq:payoff} contains the height $z_{A_f}$: $T$ can enlarge only the horizontal part of the distance, whereas $A$ can shorten the vertical part by descending.

\begin{remark}\label{rem:symmetry}
If all three players moved freely in space, the reflection in the plane through $\mathbf{x}_A$, $\mathbf{x}_D$ and $\mathbf{x}_T$ would leave the game unchanged, and the analysis would reduce to that plane. Here only horizontal translations, rotations about the vertical axis and reflections in vertical planes preserve \crefrange{eq:dynamics}{eq:M}. The reduced state space is therefore five-dimensional rather than three-dimensional, and \cref{sec:degree} shows that the optimal interception point generally leaves the plane through $A$, $D$ and $T$. If $z_{A_0}=z_{D_0}=0$ and $\theta_A=\theta_D\equiv0$, the game reduces to the planar game of~\cite{shi2026constructing}.
\end{remark}

\subsection{Dominant Regions}\label{subsec:dominant}

The TAD game contains two pursuit--evasion pairs: $A$ pursues $T$, and $D$ pursues $A$. In the first pair, the evader is bound to $\mathcal{S}$, whereas its pursuer moves in $\mathcal{H}$; in the second pair, both players move in $\mathcal{H}$. Both pairs are analyzed through dominant regions~\cite{isaacs1965differential,garcia2018cooperative,shi2026constructing}.

\begin{definition}\label{def:dominant}
A point is a dominant point of an evader if the evader can reach it before being captured by its pursuer, whatever strategy the pursuer adopts. The set of all dominant points is the dominant region of the evader.
\end{definition}

Let $\mathbf{p}=(x,y,0)$ denote a point of $\mathcal{S}$ and $\mathbf{q}=(x,y,z)$ a point of $\mathcal{H}$. In the first pair, the dominant region of $T$ is a subset of $\mathcal{S}$, denoted by $\mathcal{R}_{TE}$, and its boundary is denoted by $\mathcal{A}^R$.

\begin{lemma}\label{lem:target}
Let $\|\mathbf{x}_A-\mathbf{x}_T\|_2>R$. The dominant region of $T$ and its boundary are
\begin{equation}\label{eq:RTE}
\begin{aligned}
\mathcal{R}_{TE}&=\bigl\{\mathbf{p}\in\mathcal{S}\bigm|\|\mathbf{p}-\mathbf{x}_T\|_2<\gamma\,(\|\mathbf{p}-\mathbf{x}_A\|_2-R)\bigr\}\\
\mathcal{A}^{R}&=\bigl\{\mathbf{p}\in\mathcal{S}\bigm|\|\mathbf{p}-\mathbf{x}_T\|_2=\gamma\,(\|\mathbf{p}-\mathbf{x}_A\|_2-R)\bigr\}
\end{aligned}
\end{equation}
that is, $\mathcal{A}^R$ is the curve
\begin{equation}\label{eq:AR}
\begin{aligned}
&\sqrt{(x-x_T)^2+(y-y_T)^2}\\
&\quad=\gamma\Bigl[\sqrt{(x-x_A)^2+(y-y_A)^2+z_A^2}-R\Bigr]
\end{aligned}
\end{equation}
on $\mathcal{S}$. Moreover, every ray in $\mathcal{S}$ from $\mathbf{x}_T$ meets $\mathcal{A}^R$ exactly once, so $\mathcal{R}_{TE}$ is star-shaped with respect to $\mathbf{x}_T$.
\end{lemma}

\begin{IEEEproof}
Let
\begin{equation}\label{eq:F}
F(\mathbf{p})=\gamma\,(\|\mathbf{p}-\mathbf{x}_A\|_2-R)-\|\mathbf{p}-\mathbf{x}_T\|_2.
\end{equation}
On the ray $x=x_T+s\cos\varphi$, $y=y_T+s\sin\varphi$, $s\ge0$, it holds that $\|\mathbf{p}-\mathbf{x}_T\|_2=s$, and the Cauchy--Schwarz inequality gives
\begin{equation}\label{eq:dF}
\medmuskip=1.5mu\relax \thickmuskip=2.5mu\relax
\frac{\mathrm{d}F}{\mathrm{d}s}=\gamma\,\frac{(x-x_A)\cos\varphi+(y-y_A)\sin\varphi}{\sqrt{(x-x_A)^2+(y-y_A)^2+z_A^2}}-1
\le\gamma-1<0.
\end{equation}
Besides, $F=\gamma(\|\mathbf{x}_A-\mathbf{x}_T\|_2-R)>0$ at $s=0$, and $F\le\gamma(\|\mathbf{x}_A-\mathbf{x}_T\|_2+s-R)-s\to-\infty$ as $s\to\infty$. Hence $F$ has exactly one zero $\bar{s}(\varphi)$ on the ray, with $F>0$ for $s<\bar{s}(\varphi)$ and $F<0$ for $s>\bar{s}(\varphi)$, which proves the last claim.

Let $F(\mathbf{p})>0$, and let $T$ move straight to $\mathbf{p}$. At time $t_0+\tau$, $T$ is at the point $\mathbf{p}_\tau$ of the segment with $\|\mathbf{p}_\tau-\mathbf{x}_T\|_2=\gamma\tau$, and $F(\mathbf{p}_\tau)>0$ by the star-shapedness. Since $A$ moves at unit speed,
\begin{equation}\label{eq:Tsafe}
\|\mathbf{x}_A(t_0+\tau)-\mathbf{p}_\tau\|_2\ge\|\mathbf{x}_A-\mathbf{p}_\tau\|_2-\tau>R,
\end{equation}
where the last inequality is $F(\mathbf{p}_\tau)>0$. Thus $T$ reaches $\mathbf{p}$ uncaptured, whatever $A$ does. Conversely, let $F(\mathbf{p})\le0$. The segment from $\mathbf{x}_A$ to $\mathbf{p}$ lies in the convex set $\mathcal{H}$, and $A$, flying along it, comes within $R$ of $\mathbf{p}$ at time $t_0+\|\mathbf{p}-\mathbf{x}_A\|_2-R\le t_0+\|\mathbf{p}-\mathbf{x}_T\|_2/\gamma$, no later than $T$ can arrive. Hence $\mathbf{p}$ is not a dominant point of $T$, which proves \eqref{eq:RTE}.
\end{IEEEproof}

\begin{corollary}[see also~\cite{li2022different}]\label{cor:circle}
(i) If $R=0$, $\mathcal{A}^0$ is the circle $(x-x_c)^2+(y-y_c)^2=\rho_c^2$, where
\begin{equation}\label{eq:circle}
\begin{aligned}
x_c&=\frac{x_T-\gamma^2x_A}{1-\gamma^2},\qquad y_c=\frac{y_T-\gamma^2y_A}{1-\gamma^2}\\
\rho_c^2&=\frac{\gamma^2\bigl[(x_T-x_A)^2+(y_T-y_A)^2\bigr]}{(1-\gamma^2)^2}+\frac{\gamma^2z_A^2}{1-\gamma^2}.
\end{aligned}
\end{equation}
(ii) For fixed $x_A$, $y_A$ and $\mathbf{x}_T$, $\mathcal{R}_{TE}$ grows strictly with the height of $A$: if $0\le z_A<z_A'$, then $\mathcal{R}_{TE}(z_A)\subsetneq\mathcal{R}_{TE}(z_A')$.
\end{corollary}

\begin{IEEEproof}
(i) For $R=0$, squaring \eqref{eq:AR} and collecting terms give
\begin{equation}\label{eq:circle1}
\tightmath
\begin{aligned}
&(1-\gamma^2)(x^2+y^2)-2(x_T-\gamma^2x_A)x-2(y_T-\gamma^2y_A)y\\
&\quad+x_T^2+y_T^2-\gamma^2(x_A^2+y_A^2+z_A^2)=0.
\end{aligned}
\end{equation}
Dividing \eqref{eq:circle1} by $1-\gamma^2$ and completing the squares yield the center $(x_c,y_c)$ in \eqref{eq:circle} and
\begin{equation}\label{eq:circle2}
\tightermath
\begin{aligned}
\rho_c^2&=x_c^2+y_c^2-\frac{x_T^2+y_T^2-\gamma^2(x_A^2+y_A^2+z_A^2)}{1-\gamma^2}\\
&=\frac{(x_T-\gamma^2x_A)^2-(1-\gamma^2)(x_T^2-\gamma^2x_A^2)}{(1-\gamma^2)^2}\\
&\quad+\frac{(y_T-\gamma^2y_A)^2-(1-\gamma^2)(y_T^2-\gamma^2y_A^2)}{(1-\gamma^2)^2}+\frac{\gamma^2z_A^2}{1-\gamma^2}\\
&=\frac{\gamma^2\bigl[(x_T-x_A)^2+(y_T-y_A)^2\bigr]}{(1-\gamma^2)^2}+\frac{\gamma^2z_A^2}{1-\gamma^2}.
\end{aligned}
\end{equation}
(ii) By \eqref{eq:AR}, $\|\mathbf{p}-\mathbf{x}_A\|_2$ increases strictly with $z_A\ge0$ for every $\mathbf{p}\in\mathcal{S}$, and so does $F(\mathbf{p})$ in \eqref{eq:F}. Hence $F(\mathbf{p})>0$ at height $z_A$ implies $F(\mathbf{p})>0$ at height $z_A'$. Moreover, the points of $\mathcal{A}^R(z_A)$, where $F=0$ at height $z_A$, belong to $\mathcal{R}_{TE}(z_A')$ but not to $\mathcal{R}_{TE}(z_A)$.
\end{IEEEproof}

\begin{figure}[!t]
\centering
\includegraphics[width=\columnwidth]{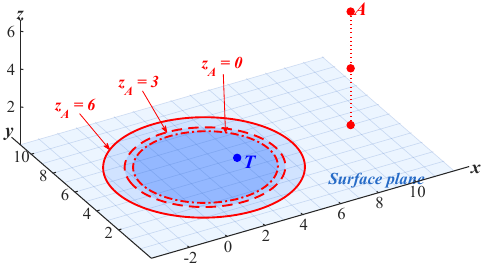}
\caption{Capture curves $\mathcal{A}^R$ formed by $T$--$A$ for $\mathbf{x}_T=(4,5,0)$, $(x_A,y_A)=(10,5)$, $\gamma=0.5$ and $R=1$. $A$ is drawn at the heights $z_A=0$, $3$ and $6$ above the same point of $\mathcal{S}$. The dominant region of $T$, enclosed by $\mathcal{A}^R$ and shaded blue (darker for lower $A$), grows with the height of $A$.}
\label{fig:capture}
\end{figure}

\begin{remark}\label{rem:capture}
$\mathcal{A}^R$ is the trace on $\mathcal{S}$ of the surface of revolution generated by a Cartesian oval with foci $\mathbf{x}_T$ and $\mathbf{x}_A$~\cite{lockwood1961book}. Indeed, squaring \eqref{eq:AR} twice removes the square roots and gives the quartic
\begin{equation}\label{eq:quartic}
\begin{aligned}
&\Bigl[(x-x_T)^2+(y-y_T)^2\\
&\quad-\gamma^2\bigl((x-x_A)^2+(y-y_A)^2+z_A^2+R^2\bigr)\Bigr]^2\\
&\quad=4\gamma^4R^2\bigl[(x-x_A)^2+(y-y_A)^2+z_A^2\bigr],
\end{aligned}
\end{equation}
of which $\mathcal{A}^R$ is the part where the bracket on the left is nonpositive and $\|\mathbf{p}-\mathbf{x}_A\|_2\ge R$. If $z_A=0$, it is the Apollonius oval of~\cite{yan2021lifeline,shi2026constructing}; if moreover $R=0$, \eqref{eq:circle} is the Apollonius circle of~\cite{garcia2018cooperative,garcia2019design}. Part~(i) of \cref{cor:circle} recovers the circle obtained in~\cite{li2022different} as the intersection of an Apollonius sphere with the plane of the evader. Part~(ii) shows that the height of $A$ acts as an extra distance: $A$ must descend before it can strike any point of $\mathcal{S}$, which enlarges the dominant region of $T$ (\cref{fig:capture}). In what follows, $\mathcal{A}^R$ is called the capture curve formed by $T$--$A$.
\end{remark}

In the second pair, the dominant region of $A$ is a subset of $\mathcal{H}$, denoted by $\mathcal{R}_{AE}$, and its boundary is denoted by $\mathcal{A}^r$. The ray from $\mathbf{x}_A$ with azimuth $\varphi$ and elevation $\psi$ consists of the points
\begin{equation}\label{eq:ray}
\begin{alignedat}{2}
x&=x_A+s\cos\psi\cos\varphi, &\qquad y&=y_A+s\cos\psi\sin\varphi\\
z&=z_A+s\sin\psi, &\qquad s&\ge0
\end{alignedat}
\end{equation}
and $\mathbf{q}_s$ denotes its point at distance $s$ from $\mathbf{x}_A$. Two quantities of the configuration recur below: the horizontal distance between $A$ and $D$, and a constant $\kappa$,
\begin{equation}\label{eq:consts}
\begin{aligned}
d_{AD}&=\sqrt{(x_A-x_D)^2+(y_A-y_D)^2}\\
\kappa&=\tfrac{1}{2}\bigl(x_D^2+y_D^2+z_D^2-x_A^2-y_A^2-z_A^2-r^2\bigr).
\end{aligned}
\end{equation}

\begin{lemma}\label{lem:attacker}
Let $\|\mathbf{x}_A-\mathbf{x}_D\|_2>r$. The dominant region of $A$ and its boundary are
\begin{equation}\label{eq:RAE}
\begin{aligned}
\mathcal{R}_{AE}&=\bigl\{\mathbf{q}\in\mathcal{H}\bigm|\|\mathbf{q}-\mathbf{x}_A\|_2<\|\mathbf{q}-\mathbf{x}_D\|_2-r\bigr\}\\
\mathcal{A}^{r}&=\bigl\{\mathbf{q}\in\mathcal{H}\bigm|\|\mathbf{q}-\mathbf{x}_A\|_2=\|\mathbf{q}-\mathbf{x}_D\|_2-r\bigr\}
\end{aligned}
\end{equation}
and $\mathcal{R}_{AE}$ is convex. Moreover, the ray \eqref{eq:ray} meets the surface $\|\mathbf{q}-\mathbf{x}_A\|_2=\|\mathbf{q}-\mathbf{x}_D\|_2-r$ if and only if $\varpi(\varphi,\psi)>0$, and then only at $s=\rho(\varphi,\psi)$, where
\begin{equation}\label{eq:rho}
\tightmath
\begin{aligned}
\rho(\varphi,\psi)&=\frac{d_{AD}^2+(z_A-z_D)^2-r^2}{2\,\varpi(\varphi,\psi)}\\
\varpi(\varphi,\psi)&=r-(x_A-x_D)\cos\psi\cos\varphi\\
&\quad-(y_A-y_D)\cos\psi\sin\varphi-(z_A-z_D)\sin\psi.
\end{aligned}
\end{equation}
\end{lemma}

\begin{IEEEproof}
Let
\begin{equation}\label{eq:G}
G(\mathbf{q})=\|\mathbf{q}-\mathbf{x}_D\|_2-r-\|\mathbf{q}-\mathbf{x}_A\|_2.
\end{equation}
Along the ray \eqref{eq:ray}, $\|\mathbf{q}_s-\mathbf{x}_A\|_2=s$, and the Cauchy--Schwarz inequality gives, for $G=G(\mathbf{q}_s)$,
\begin{equation}\label{eq:dG}
\begin{aligned}
\frac{\mathrm{d}G}{\mathrm{d}s}&=\bigl[(x-x_D)\cos\psi\cos\varphi+(y-y_D)\cos\psi\sin\varphi\\
&\quad+(z-z_D)\sin\psi\bigr]\frac{1}{\|\mathbf{q}_s-\mathbf{x}_D\|_2}-1\le0.
\end{aligned}
\end{equation}
Let $G(\mathbf{q})>0$, and let $A$ fly straight to $\mathbf{q}$, that is, along the ray \eqref{eq:ray} whose azimuth and elevation point from $\mathbf{x}_A$ to $\mathbf{q}$; the segment lies in $\mathcal{H}$ because $\mathcal{H}$ is convex. By \eqref{eq:dG}, $G(\mathbf{q}_s)\ge G(\mathbf{q})>0$ for $0\le s\le\|\mathbf{q}-\mathbf{x}_A\|_2$. At time $t_0+s$, $A$ is at $\mathbf{q}_s$, and $D$, moving at unit speed, satisfies
\begin{equation}\label{eq:Asafe}
\medmuskip=2mu\relax \thickmuskip=3mu\relax
\|\mathbf{x}_D(t_0+s)-\mathbf{q}_s\|_2\ge\|\mathbf{x}_D-\mathbf{q}_s\|_2-s
=G(\mathbf{q}_s)+r>r.
\end{equation}
Thus $A$ reaches $\mathbf{q}$ unintercepted, whatever $D$ does. Conversely, if $G(\mathbf{q})\le0$, $D$ flies straight to $\mathbf{q}$ inside $\mathcal{H}$ and comes within $r$ of $\mathbf{q}$ at time $t_0+\|\mathbf{q}-\mathbf{x}_D\|_2-r\le t_0+\|\mathbf{q}-\mathbf{x}_A\|_2$. This proves \eqref{eq:RAE}.

Since both sides are nonnegative, $\|\mathbf{q}-\mathbf{x}_A\|_2+r<\|\mathbf{q}-\mathbf{x}_D\|_2$ holds if and only if $(\|\mathbf{q}-\mathbf{x}_A\|_2+r)^2<\|\mathbf{q}-\mathbf{x}_D\|_2^2$, that is, $2r\|\mathbf{q}-\mathbf{x}_A\|_2<\|\mathbf{q}-\mathbf{x}_D\|_2^2-\|\mathbf{q}-\mathbf{x}_A\|_2^2-r^2$. By \eqref{eq:consts}, the right-hand side is affine in $\mathbf{q}$:
\begin{equation}\label{eq:affine0}
\tightmath
\begin{aligned}
&\|\mathbf{q}-\mathbf{x}_D\|_2^2-\|\mathbf{q}-\mathbf{x}_A\|_2^2-r^2\\
&\quad=(x-x_D)^2+(y-y_D)^2+(z-z_D)^2\\
&\qquad-(x-x_A)^2-(y-y_A)^2-(z-z_A)^2-r^2\\
&\quad=2\bigl[(x_A-x_D)x+(y_A-y_D)y+(z_A-z_D)z+\kappa\bigr].
\end{aligned}
\end{equation}
Halving the inequality therefore gives
\begin{equation}\label{eq:affine}
\mbox{\small$\medmuskip=2mu\relax \thickmuskip=3mu\relax
r\|\mathbf{q}-\mathbf{x}_A\|_2<(x_A-x_D)x+(y_A-y_D)y
+(z_A-z_D)z+\kappa.$}
\end{equation}
The left-hand side of \eqref{eq:affine} is convex in $\mathbf{q}$, and the right-hand side is affine. Hence \eqref{eq:affine} defines a convex set, and so does its intersection $\mathcal{R}_{AE}$ with the convex set $\mathcal{H}$.

Finally, substituting \eqref{eq:ray} into $\|\mathbf{q}_s-\mathbf{x}_D\|_2=s+r$ and squaring both sides give
\begin{equation}\label{eq:rho1}
\begin{aligned}
&(x_A-x_D+s\cos\psi\cos\varphi)^2\\
&\quad+(y_A-y_D+s\cos\psi\sin\varphi)^2\\
&\quad+(z_A-z_D+s\sin\psi)^2=(s+r)^2.
\end{aligned}
\end{equation}
Expanding the squares in \eqref{eq:rho1} gives
\begin{equation}\label{eq:rho15}
\tightermath
\begin{aligned}
&(x_A-x_D)^2+(y_A-y_D)^2+(z_A-z_D)^2+s^2\\
&\quad+2s\bigl[(x_A-x_D)\cos\psi\cos\varphi+(y_A-y_D)\cos\psi\sin\varphi\\
&\quad+(z_A-z_D)\sin\psi\bigr]=s^2+2rs+r^2.
\end{aligned}
\end{equation}
The terms in $s^2$ cancel, and by \eqref{eq:consts} and \eqref{eq:rho}, \eqref{eq:rho15} becomes
\begin{equation}\label{eq:rho2}
2s\,\varpi(\varphi,\psi)=d_{AD}^2+(z_A-z_D)^2-r^2,
\end{equation}
whose right-hand side equals $\|\mathbf{x}_A-\mathbf{x}_D\|_2^2-r^2>0$. As $s+r>0$, every nonnegative root of \eqref{eq:rho2} satisfies $\|\mathbf{q}_s-\mathbf{x}_D\|_2=s+r$. Hence a root $s>0$ exists if and only if $\varpi(\varphi,\psi)>0$, and it is $\rho(\varphi,\psi)$ in \eqref{eq:rho}.
\end{IEEEproof}

\begin{remark}\label{rem:interception}
$\mathcal{A}^r$ is the part in $\mathcal{H}$ of the sheet nearer $\mathbf{x}_A$ of a hyperboloid of revolution of two sheets with foci $\mathbf{x}_A$ and $\mathbf{x}_D$; in the plane, it reduces to the interception hyperbola of~\cite{garcia2018cooperative,shi2026constructing}. We call $\mathcal{A}^r$ the interception surface formed by $D$--$A$. Formula \eqref{eq:rho} describes $\mathcal{A}^r$ as seen from $A$: if $A$ flies straight with azimuth $\varphi$ and elevation $\psi$, $D$ can intercept it no earlier than $t_0+\rho(\varphi,\psi)$, and attains this time by flying straight to $\mathbf{q}_{\rho(\varphi,\psi)}$. If $\varpi(\varphi,\psi)\le0$, $D$ cannot intercept $A$ on this ray at all, although for $\psi<0$ the ray leaves $\mathcal{H}$ at $s=-z_A/\sin\psi$.
\end{remark}

The trace $\mathcal{C}^r=\mathcal{A}^r\cap\mathcal{S}$ bounds the part of the surface that $A$ can reach before interception. Let
\begin{equation}\label{eq:ell}
\ell(x,y)=(x_A-x_D)\,x+(y_A-y_D)\,y+\kappa,
\end{equation}
which is the right-hand side of \eqref{eq:affine} on $\mathcal{S}$.

\begin{lemma}\label{lem:trace}
Let $\|\mathbf{x}_A-\mathbf{x}_D\|_2>r$. The trace of the interception surface on $\mathcal{S}$ is
\begin{equation}\label{eq:trace}
\begin{aligned}
\Gamma^r&=\bigl\{\mathbf{p}\in\mathcal{S}\bigm|\ell^2(x,y)=r^2\|\mathbf{p}-\mathbf{x}_A\|_2^2\bigr\}\\
\mathcal{C}^r&=\bigl\{\mathbf{p}\in\Gamma^r\bigm|\ell(x,y)\ge0\bigr\}
\end{aligned}
\end{equation}
where $\Gamma^r$ is a conic with discriminant
\begin{equation}\label{eq:Delta}
\Delta=4r^2\bigl(d_{AD}^2-r^2\bigr).
\end{equation}
If $\mathcal{R}_{AE}\cap\mathcal{S}\neq\emptyset$, $\mathcal{C}^r$ is the boundary of $\mathcal{R}_{AE}\cap\mathcal{S}$ in $\mathcal{S}$, and it is one branch of a hyperbola, a parabola or an ellipse according as $d_{AD}>r$, $d_{AD}=r$ or $d_{AD}<r$; in the last case, $\mathcal{R}_{AE}\cap\mathcal{S}$ is bounded. If $\mathcal{R}_{AE}\cap\mathcal{S}=\emptyset$, $\mathcal{C}^r$ has at most one point, and this occurs only if $d_{AD}\le r$.
\end{lemma}

\begin{IEEEproof}
On $\mathcal{S}$, $z=0$, and the equality case of \eqref{eq:affine} reads $r\|\mathbf{p}-\mathbf{x}_A\|_2=\ell(x,y)$. It holds if and only if $\ell(x,y)\ge0$ and $\ell^2(x,y)=r^2\|\mathbf{p}-\mathbf{x}_A\|_2^2$, which proves \eqref{eq:trace}. Expanding $\ell^2(x,y)-r^2\|\mathbf{p}-\mathbf{x}_A\|_2^2$ by \eqref{eq:ell}, $\Gamma^r$ is the conic
\begin{equation}\label{eq:conic}
\varepsilon_{11}x^2+2\varepsilon_{12}xy+\varepsilon_{22}y^2+2\varepsilon_{13}x+2\varepsilon_{23}y+\varepsilon_{33}=0,
\end{equation}
whose coefficients are
\begin{equation}\label{eq:quad}
\begin{gathered}
\varepsilon_{11}=(x_A-x_D)^2-r^2,\qquad \varepsilon_{22}=(y_A-y_D)^2-r^2\\
(\varepsilon_{13},\varepsilon_{23})=\kappa\,(x_A-x_D,\,y_A-y_D)+r^2(x_A,\,y_A)\\
\varepsilon_{12}=(x_A-x_D)(y_A-y_D),\qquad \varepsilon_{33}=\kappa^2-r^2\|\mathbf{x}_A\|_2^2
\end{gathered}
\end{equation}
so that $\Delta=4(\varepsilon_{12}^2-\varepsilon_{11}\varepsilon_{22})$ and
\begin{equation}\label{eq:Delta1}
\begin{aligned}
\Delta&=4(x_A-x_D)^2(y_A-y_D)^2\\
&\quad-4\bigl[(x_A-x_D)^2-r^2\bigr]\bigl[(y_A-y_D)^2-r^2\bigr]\\
&=4r^2\bigl[(x_A-x_D)^2+(y_A-y_D)^2-r^2\bigr]\\
&=4r^2\bigl(d_{AD}^2-r^2\bigr).
\end{aligned}
\end{equation}
$\Gamma^r$ is the trace on $\mathcal{S}$ of the whole hyperboloid $\bigl|\|\mathbf{q}-\mathbf{x}_D\|_2-\|\mathbf{q}-\mathbf{x}_A\|_2\bigr|=r$, which contains no straight line. Hence $\Gamma^r$ contains no segment, and when it has more than one point, its type is given by the sign of $\Delta$~\cite{audin2003geometry}.

Next, $\ell\neq0$ on $\Gamma^r$. Indeed, $\ell=0$ on $\Gamma^r$ forces $\|\mathbf{p}-\mathbf{x}_A\|_2=0$, that is, $\mathbf{p}=\mathbf{x}_A$ and $z_A=0$; but then \eqref{eq:consts} and \eqref{eq:ell} give $\ell(x_A,y_A)=\tfrac12(\|\mathbf{x}_A-\mathbf{x}_D\|_2^2-r^2)>0$. Hence $\ell$ keeps its sign on each connected component of $\Gamma^r$, and $\mathcal{C}^r$ is a union of such components.

Let $f(\mathbf{p})=r\|\mathbf{p}-\mathbf{x}_A\|_2-\ell(x,y)$, which is convex on $\mathcal{S}$, so that $\mathcal{R}_{AE}\cap\mathcal{S}=\{f<0\}$ and $\mathcal{C}^r=\{f=0\}$ by \eqref{eq:affine}. Suppose that $f(\mathbf{p}_1)<0$ for some $\mathbf{p}_1$. If $f(\mathbf{p}_0)=0$, convexity gives $f(\mathbf{p}_0+\tau(\mathbf{p}_1-\mathbf{p}_0))\le\tau f(\mathbf{p}_1)<0$ for $\tau\in(0,1]$, so every point of $\mathcal{C}^r$ is a boundary point of $\{f<0\}$; the converse follows from continuity. An ellipse and a parabola are connected, so $\mathcal{C}^r$ is the whole curve. A hyperbola has two branches, and the convex set $\{f<0\}$ cannot be bounded by both, because the region between the branches is not convex; so $\mathcal{C}^r$ is one branch. If $d_{AD}<r$, $\mathcal{C}^r$ is an ellipse, and $\{f<0\}$, the convex region it encloses, is bounded.

Suppose instead that $f\ge0$ on $\mathcal{S}$. Then $\mathcal{C}^r=\{f\le0\}$ is convex and lies in $\Gamma^r$; as $\Gamma^r$ contains no segment, $\mathcal{C}^r$ has at most one point. Finally, let $d_{AD}>r$, and choose a horizontal unit vector $(u_x,u_y)$ with $(x_A-x_D)u_x+(y_A-y_D)u_y>r$. Along $\mathbf{p}=(x_A+su_x,y_A+su_y,0)$,
\begin{equation}\label{eq:Glimit}
\begin{aligned}
\lim_{s\to\infty}G(\mathbf{p})&=\lim_{s\to\infty}\frac{2\ell(x,y)+r^2}{\|\mathbf{p}-\mathbf{x}_D\|_2+\|\mathbf{p}-\mathbf{x}_A\|_2}-r\\
&=(x_A-x_D)u_x+(y_A-y_D)u_y-r>0,
\end{aligned}
\end{equation}
so $\mathcal{R}_{AE}\cap\mathcal{S}\neq\emptyset$. This completes the proof.
\end{IEEEproof}

\begin{figure}[!t]
\centering
\includegraphics[width=\columnwidth]{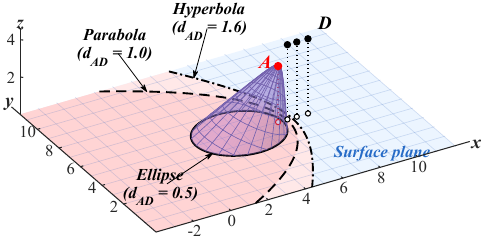}
\caption{Traces $\mathcal{C}^r$ of the interception surface formed by $D$--$A$ for $\mathbf{x}_A=(6,5,3)$, $z_D=4$ and $r=1$. $D$ is drawn at the horizontal distances $d_{AD}=0.5$, $1$ and $1.6$ from $A$, and the trace is an ellipse, a parabola and one branch of a hyperbola; the dominant region of $A$ on $\mathcal{S}$ inside it is shaded red (darker for smaller $d_{AD}$). The translucent sheet is the interception surface $\mathcal{A}^r$ for $d_{AD}=0.5$; it hangs below $A$ and meets $\mathcal{S}$ in the ellipse.}
\label{fig:traces}
\end{figure}

\begin{remark}\label{rem:topology}
The type of $\mathcal{C}^r$ depends only on $d_{AD}$, not on the heights $z_A$ and $z_D$ (\cref{fig:traces}). In the planar game, $d_{AD}=\|\mathbf{x}_A-\mathbf{x}_D\|_2>r$, so the interception curve is always a hyperbola. Here $d_{AD}<r$ whenever $D$ is nearly above or below $A$, and the surface points that $A$ can reach before interception then form a bounded set. If $D$ is directly below $A$, this set is empty, since $\|\mathbf{p}-\mathbf{x}_D\|_2<\|\mathbf{p}-\mathbf{x}_A\|_2$ for every $\mathbf{p}\in\mathcal{S}$; $D$ then shields the whole surface. This topological change has no planar counterpart and underlies the bounded winning region of $A$ in \cref{sec:kind}.
\end{remark}

\Crefrange{lem:target}{lem:trace} describe the two pursuit--evasion pairs separately. \cref{sec:degree} couples them in the game of degree on the interception surface \eqref{eq:rho}, and \cref{sec:kind} brings in the capture radius $R$ to solve the game of kind.

\section{Game of Degree}\label{sec:degree}

Existing TAD studies~\cite{garcia2018cooperative,garcia2019design,liang2019differential,shi2026constructing} construct the barrier of the game of kind first, and then solve the game of degree inside the winning region of the $D$--$T$ team. For $R>0$, the barrier of~\cite{shi2025determination,shi2026constructing} is built from straight strikes in which $A$, its interception point and the capture point are collinear. Such a strike is not optimal for $A$: a capture only requires the distance $R$, so $A$ gains by veering away from $D$ before it strikes, and the barrier so constructed is not semipermeable (\cref{sec:kind}). This article therefore reverses the order. The game of degree is solved first, and its value function yields the barrier in \cref{sec:kind}. To this end, $\mathcal{M}_1$ is ignored and the game is terminated at interception, so that $V$ in \eqref{eq:value} is defined for every state. Moreover, unlike the planar game, the interception point lies in the air or on the surface, and $T$ either flees from its foot or runs under it. \Cref{subsec:straight} derives the straight-line extremals and solves the evasion in the air and the solutions on the surface; \cref{subsec:overhead} shows that a strike directly above $T$ is not optimal and solves the crossing in the air.

\subsection{Straight-Line Solutions}\label{subsec:straight}

Let $\bm{\lambda}=(\lambda_{x_A},\lambda_{y_A},\lambda_{z_A},\lambda_{x_D},\lambda_{y_D},\lambda_{z_D},\lambda_{x_T},\lambda_{y_T})$ denote the costate of \eqref{eq:dynamics}. The Hamiltonian of the game is
\begin{equation}\label{eq:H}
\tightmath
\begin{aligned}
\mathscr{H}&=\lambda_{x_A}\cos\theta_A\cos\beta+\lambda_{y_A}\cos\theta_A\sin\beta+\lambda_{z_A}\sin\theta_A\\
&\quad+\lambda_{x_D}\cos\theta_D\cos\delta+\lambda_{y_D}\cos\theta_D\sin\delta\\
&\quad+\lambda_{z_D}\sin\theta_D+\gamma\lambda_{x_T}\cos\zeta+\gamma\lambda_{y_T}\sin\zeta.
\end{aligned}
\end{equation}
For a point $Q=(x,y,z)\in\mathcal{H}$, the distances to $A$, to $D$ and, horizontally, to $T$ are
\begin{equation}\label{eq:dQ}
\begin{aligned}
d_A&=\sqrt{(x-x_A)^2+(y-y_A)^2+(z-z_A)^2}\\
d_D&=\sqrt{(x-x_D)^2+(y-y_D)^2+(z-z_D)^2}\\
d_T&=\sqrt{(x-x_T)^2+(y-y_T)^2}
\end{aligned}
\end{equation}
and the corresponding direction cosines are
\begin{equation}\label{eq:cosines}
\begin{alignedat}{3}
a_x&=\frac{x-x_A}{d_A}, &\quad a_y&=\frac{y-y_A}{d_A}, &\quad a_z&=\frac{z-z_A}{d_A}\\
b_x&=\frac{x-x_D}{d_D}, &\quad b_y&=\frac{y-y_D}{d_D}, &\quad b_z&=\frac{z-z_D}{d_D}\\
c_x&=\frac{x-x_T}{d_T}, &\quad c_y&=\frac{y-y_T}{d_T}. & &
\end{alignedat}
\end{equation}
Two payoffs of the point $Q$ appear below, the evasion payoff $J$ and the crossing payoff $K$:
\begin{equation}\label{eq:JK}
\tightermath
J(Q)=\sqrt{u^2+z^2},\; u=d_T+\gamma d_A,\;
K(Q)=\gamma d_A-d_T.
\end{equation}

\begin{lemma}\label{lem:extremal}
For the game \crefrange{eq:dynamics}{eq:value}, terminated at interception, all controls are constant along an extremal of the two-sided maximum principle. $A$ and $D$ move straight to a common point $Q=(x,y,z)\in\mathcal{A}^r$, which $A$ reaches at $t_f$ while $D$ reaches the distance $r$ from it. $T$ moves straight on $\mathcal{S}$, and its velocity at $t_f$ points away from $(x,y)$. If $T$ moves away from $(x,y)$ from the start, the payoff is $J(Q)$; if $T$ moves toward $(x,y)$ and passes it before $t_f$, the payoff is $\sqrt{K^2(Q)+z^2}$.
\end{lemma}

\begin{IEEEproof}
Isaacs' main equation $\min_{\beta,\theta_A}\max_{\delta,\theta_D,\zeta}\mathscr{H}=0$~\cite{isaacs1965differential,basar1998dynamic} requires the stationarity of \eqref{eq:H} in each control, so that the derivatives
\begin{equation}\label{eq:Hstat}
\begin{aligned}
\frac{\partial\mathscr{H}}{\partial\beta}&=\cos\theta_A\bigl(\lambda_{y_A}\cos\beta-\lambda_{x_A}\sin\beta\bigr)\\
\frac{\partial\mathscr{H}}{\partial\theta_A}&=\lambda_{z_A}\cos\theta_A-\sin\theta_A\bigl(\lambda_{x_A}\cos\beta+\lambda_{y_A}\sin\beta\bigr)\\
\frac{\partial\mathscr{H}}{\partial\delta}&=\cos\theta_D\bigl(\lambda_{y_D}\cos\delta-\lambda_{x_D}\sin\delta\bigr)\\
\frac{\partial\mathscr{H}}{\partial\theta_D}&=\lambda_{z_D}\cos\theta_D-\sin\theta_D\bigl(\lambda_{x_D}\cos\delta+\lambda_{y_D}\sin\delta\bigr)\\
\frac{\partial\mathscr{H}}{\partial\zeta}&=\gamma\bigl(\lambda_{y_T}\cos\zeta-\lambda_{x_T}\sin\zeta\bigr)
\end{aligned}
\end{equation}
all vanish. Let $\mathbf{v}_A=(\cos\theta_A\cos\beta,\cos\theta_A\sin\beta,\sin\theta_A)$, $\mathbf{v}_D$ be defined likewise and $\mathbf{v}_T=(\cos\zeta,\sin\zeta)$. Each pair in \eqref{eq:Hstat} makes a unit velocity parallel to the costate of its player, and since $A$ minimizes while $D$ and $T$ maximize, the Cauchy--Schwarz inequality gives
\begin{equation}\label{eq:Hminmax}
\begin{aligned}
\min_{\beta,\theta_A}\bm{\lambda}_A\cdot\mathbf{v}_A&=-\|\bm{\lambda}_A\|_2\\
\max_{\delta,\theta_D}\bm{\lambda}_D\cdot\mathbf{v}_D&=\|\bm{\lambda}_D\|_2\\
\max_{\zeta}\gamma\bm{\lambda}_T\cdot\mathbf{v}_T&=\gamma\|\bm{\lambda}_T\|_2,
\end{aligned}
\end{equation}
attained by the optimal controls
\begin{equation}\label{eq:uopt}
\begin{alignedat}{2}
\cos\theta_A^*\cos\beta^*&=-\frac{\lambda_{x_A}}{\|\bm{\lambda}_A\|_2}, &\quad \cos\theta_D^*\cos\delta^*&=\frac{\lambda_{x_D}}{\|\bm{\lambda}_D\|_2}\\
\cos\theta_A^*\sin\beta^*&=-\frac{\lambda_{y_A}}{\|\bm{\lambda}_A\|_2}, &\quad \cos\theta_D^*\sin\delta^*&=\frac{\lambda_{y_D}}{\|\bm{\lambda}_D\|_2}\\
\sin\theta_A^*&=-\frac{\lambda_{z_A}}{\|\bm{\lambda}_A\|_2}, &\quad \sin\theta_D^*&=\frac{\lambda_{z_D}}{\|\bm{\lambda}_D\|_2}\\
\cos\zeta^*&=\frac{\lambda_{x_T}}{\|\bm{\lambda}_T\|_2}, &\quad \sin\zeta^*&=\frac{\lambda_{y_T}}{\|\bm{\lambda}_T\|_2}
\end{alignedat}
\end{equation}
where $\bm{\lambda}_A=(\lambda_{x_A},\lambda_{y_A},\lambda_{z_A})$, $\bm{\lambda}_D=(\lambda_{x_D},\lambda_{y_D},\lambda_{z_D})$ and $\bm{\lambda}_T=(\lambda_{x_T},\lambda_{y_T})$. Since $\mathscr{H}$ does not depend on the state, the costate equations read
\begin{equation}\label{eq:costate}
\begin{alignedat}{2}
\dot{\lambda}_{x_A}&=-\frac{\partial\mathscr{H}}{\partial x_A}=0, &\qquad \dot{\lambda}_{x_D}&=-\frac{\partial\mathscr{H}}{\partial x_D}=0\\
\dot{\lambda}_{y_A}&=-\frac{\partial\mathscr{H}}{\partial y_A}=0, &\qquad \dot{\lambda}_{y_D}&=-\frac{\partial\mathscr{H}}{\partial y_D}=0\\
\dot{\lambda}_{z_A}&=-\frac{\partial\mathscr{H}}{\partial z_A}=0, &\qquad \dot{\lambda}_{z_D}&=-\frac{\partial\mathscr{H}}{\partial z_D}=0\\
\dot{\lambda}_{x_T}&=-\frac{\partial\mathscr{H}}{\partial x_T}=0, &\qquad \dot{\lambda}_{y_T}&=-\frac{\partial\mathscr{H}}{\partial y_T}=0,
\end{alignedat}
\end{equation}
so $\bm{\lambda}$ and the controls \eqref{eq:uopt} are constant, and all players move straight. A segment between two points of the convex set $\mathcal{H}$ stays in $\mathcal{H}$; hence \eqref{eq:constraint} is inactive on $[t_0,t_f)$, and it can only be active at $t_f$, where it adds the term $-\eta$, with a multiplier $\eta\ge0$, to $\lambda_{z_A}$~\cite{hartl1995survey}.

At $t_f$, $A$ is intercepted, so that $\|\mathbf{x}_{A_f}-\mathbf{x}_{D_f}\|_2=r$. With a Lagrange multiplier $\nu$, the augmented terminal function is
\begin{equation}\label{eq:Phia}
\Phi_a(\mathbf{x}_f)=\|\mathbf{x}_{A_f}-\mathbf{x}_{T_f}\|_2+\nu\bigl(\|\mathbf{x}_{A_f}-\mathbf{x}_{D_f}\|_2-r\bigr),
\end{equation}
and the transversality condition $\bm{\lambda}(t_f)=\partial\Phi_a/\partial\mathbf{x}_f$ gives, with $\Phi_f=\|\mathbf{x}_{A_f}-\mathbf{x}_{T_f}\|_2$,
\begin{equation}\label{eq:transA}
\begin{aligned}
\lambda_{x_A}&=\frac{x_{A_f}-x_{T_f}}{\Phi_f}+\nu\,\frac{x_{A_f}-x_{D_f}}{r}\\
\lambda_{y_A}&=\frac{y_{A_f}-y_{T_f}}{\Phi_f}+\nu\,\frac{y_{A_f}-y_{D_f}}{r}\\
\lambda_{z_A}&=\frac{z_{A_f}}{\Phi_f}+\nu\,\frac{z_{A_f}-z_{D_f}}{r}-\eta
\end{aligned}
\end{equation}
\begin{equation}\label{eq:transD}
\begin{aligned}
\lambda_{x_D}&=-\nu\,\frac{x_{A_f}-x_{D_f}}{r}\\
\lambda_{y_D}&=-\nu\,\frac{y_{A_f}-y_{D_f}}{r}\\
\lambda_{z_D}&=-\nu\,\frac{z_{A_f}-z_{D_f}}{r}
\end{aligned}
\end{equation}
\begin{equation}\label{eq:transT}
\lambda_{x_T}=-\frac{x_{A_f}-x_{T_f}}{\Phi_f},\qquad
\lambda_{y_T}=-\frac{y_{A_f}-y_{T_f}}{\Phi_f}.
\end{equation}
By \eqref{eq:uopt} and \eqref{eq:transT}, the velocity of $T$ points from $(x_{A_f},y_{A_f})$ to $(x_{T_f},y_{T_f})$. By \eqref{eq:uopt} and \eqref{eq:transD}, the velocity of $D$ is parallel to $\mathbf{x}_{A_f}-\mathbf{x}_{D_f}$; since $D$ moves straight and ends at the distance $r$ from $Q=\mathbf{x}_{A_f}$, it flies toward $Q$ and $t_f-t_0=d_D-r$. $A$ flies straight to $Q$, so $t_f-t_0=d_A$, and $d_A=d_D-r$, that is, $Q\in\mathcal{A}^r$. With \eqref{eq:cosines}, the terminal positions of $A$ and $D$ are
\begin{equation}\label{eq:xf}
\begin{alignedat}{3}
x_{A_f}&=x, &\quad y_{A_f}&=y, &\quad z_{A_f}&=z\\
x_{D_f}&=x-rb_x, &\quad y_{D_f}&=y-rb_y, &\quad z_{D_f}&=z-rb_z.
\end{alignedat}
\end{equation}
As $T$ moves straight and its final velocity points away from $(x,y)$, the point $(x,y)$ lies on the line of motion of $T$. If $T$ moves away from $(x,y)$ from the start, it covers the distance $\gamma d_A$ and ends at
\begin{equation}\label{eq:Tf}
x_{T_f}=x_T-\gamma d_A c_x,\qquad y_{T_f}=y_T-\gamma d_A c_y,
\end{equation}
and if $T$ passes through $(x,y)$, the signs in \eqref{eq:Tf} are reversed. Hence
\begin{equation}\label{eq:Phif}
\Phi_f=\begin{cases}
\sqrt{(d_T+\gamma d_A)^2+z^2}, & \text{if $T$ flees from $(x,y)$}\\[2pt]
\sqrt{(\gamma d_A-d_T)^2+z^2}, & \text{if $T$ crosses $(x,y)$}
\end{cases}
\end{equation}
which are $J(Q)$ and $\sqrt{K^2(Q)+z^2}$ in \eqref{eq:JK}.
\end{IEEEproof}

\Cref{lem:extremal} leaves the order of commitment open. If $A$ commits to $Q$, $T$ flees from its foot, and the payoff is $J(Q)$. If $T$ commits to its heading $\zeta$, $A$ and $D$ reply with straight flights to a point $Q$ of $\mathcal{A}^r$, and the payoff is
\begin{equation}\label{eq:Fz}
F_\zeta(Q)=\bigl\|Q-\mathbf{x}_T-\gamma d_A(\cos\zeta,\sin\zeta,0)\bigr\|_2,
\end{equation}
unless $A$ meets $T$ inside $\mathcal{R}_{AE}\cap\mathcal{S}$, which yields zero. Let $\zeta_Q$ be the heading from $\mathbf{x}_T$ to the foot of $Q$, so that $(\cos\zeta_Q,\sin\zeta_Q)=(c_x,c_y)$. In coordinates,
\begin{equation}\label{eq:Fz2}
\begin{aligned}
F_\zeta^2(Q)&=(x-x_T-\gamma d_A\cos\zeta)^2\\
&\quad+(y-y_T-\gamma d_A\sin\zeta)^2+z^2\\
&=d_T^2+\gamma^2d_A^2+z^2-2\gamma d_Ad_T\cos(\zeta-\zeta_Q),
\end{aligned}
\end{equation}
so that $\min_\zeta F_\zeta(Q)=\sqrt{K^2(Q)+z^2}$, attained when $T$ runs toward the foot of $Q$, and $\max_\zeta F_\zeta(Q)=J(Q)$, attained when $T$ runs away from it. $T$ guarantees the lower value and $A$ the upper value, so that
\begin{equation}\label{eq:Vpm}
V^-=\max_{\zeta}\,\min_{Q\in\mathcal{A}^r}F_\zeta(Q)\le V\le\min_{Q\in\mathcal{A}^r}J(Q)=V^+.
\end{equation}
The theorems below identify $V$ where the two bounds meet, and where $V^-$ itself satisfies the HJI equation.

\begin{theorem}[Evasion in the air]\label{thm:air}
For the game \crefrange{eq:dynamics}{eq:value}, under the condition that $\|\mathbf{x}_T-\mathbf{x}_A\|_2>\|\mathbf{x}_T-\mathbf{x}_D\|_2-r$, let $Q^*=(x^*,y^*,z^*)$ be a nondegenerate minimizer of $J$ over $\mathcal{A}^r$ with $z^*>0$ and $(x^*,y^*)\neq(x_T,y_T)$. If the multiplier $\mu$ in \eqref{eq:stat} is positive, the optimal strategies of $A$, $D$ and $T$ are
\begin{equation}\label{eq:strat}
\begin{alignedat}{2}
\cos\theta_A^*\cos\beta^*&=\frac{x^*-x_A}{d_A}, &\quad \cos\theta_D^*\cos\delta^*&=\frac{x^*-x_D}{d_D}\\
\cos\theta_A^*\sin\beta^*&=\frac{y^*-y_A}{d_A}, &\quad \cos\theta_D^*\sin\delta^*&=\frac{y^*-y_D}{d_D}\\
\sin\theta_A^*&=\frac{z^*-z_A}{d_A}, &\quad \sin\theta_D^*&=\frac{z^*-z_D}{d_D}\\
\cos\zeta^*&=\frac{x_T-x^*}{d_T}, &\quad \sin\zeta^*&=\frac{y_T-y^*}{d_T}
\end{alignedat}
\end{equation}
where \eqref{eq:dQ} is evaluated at $Q^*$, and $Q^*$ and $\mu$ satisfy $d_A=d_D-r$ and
\begin{equation}\label{eq:stat}
\begin{aligned}
(d_T+\gamma d_A)(c_x+\gamma a_x)&=\mu J(b_x-a_x)\\
(d_T+\gamma d_A)(c_y+\gamma a_y)&=\mu J(b_y-a_y)\\
\gamma(d_T+\gamma d_A)a_z+z^*&=\mu J(b_z-a_z).
\end{aligned}
\end{equation}
The value function
\begin{equation}\label{eq:Vair}
V(\mathbf{x})=J(Q^*)=\sqrt{(d_T+\gamma d_A)^2+z^{*2}}
\end{equation}
is $C^1$ and satisfies the HJI equation
\begin{equation}\label{eq:HJI}
-\frac{\partial V}{\partial t}=\frac{\partial V}{\partial\mathbf{x}}\cdot\mathbf{f}(\mathbf{x},\beta^*,\theta_A^*,\delta^*,\theta_D^*,\zeta^*)=0.
\end{equation}
\end{theorem}

\begin{IEEEproof}
By \cref{lem:extremal}, an extremal along which $T$ moves away from $(x,y)\neq(x_T,y_T)$ is fixed by its interception point $Q\in\mathcal{A}^r$, and its payoff is $J(Q)$. The strategies \eqref{eq:strat} are the straight motions of \cref{lem:extremal} toward $Q^*$ and away from $(x^*,y^*)$. Since $\mathscr{H}$ does not depend on $t$ and $t_f$ is free, $\mathscr{H}\equiv0$ along the extremal, and substituting \eqref{eq:uopt} into \eqref{eq:H} gives
\begin{equation}\label{eq:H0}
-\|\bm{\lambda}_A\|_2+\|\bm{\lambda}_D\|_2+\gamma\|\bm{\lambda}_T\|_2=0.
\end{equation}
Let $\mu=-\nu$ and $u=d_T+\gamma d_A$. Since $x^*-x_T=d_Tc_x$ and $y^*-y_T=d_Tc_y$, \eqref{eq:xf} and \eqref{eq:Tf} give the terminal differences
\begin{equation}\label{eq:diffs}
\begin{alignedat}{2}
x_{A_f}-x_{T_f}&=uc_x, &\qquad x_{A_f}-x_{D_f}&=rb_x\\
y_{A_f}-y_{T_f}&=uc_y, &\qquad y_{A_f}-y_{D_f}&=rb_y\\
z_{A_f}-z_{T_f}&=z^*, &\qquad z_{A_f}-z_{D_f}&=rb_z
\end{alignedat}
\end{equation}
with $\Phi_f=J=J(Q^*)$, and \crefrange{eq:transA}{eq:transT} become
\begin{equation}\label{eq:lamQ}
\begin{alignedat}{2}
\lambda_{x_A}&=\frac{uc_x}{J}-\mu b_x, &\qquad \lambda_{x_D}&=\mu b_x\\
\lambda_{y_A}&=\frac{uc_y}{J}-\mu b_y, &\qquad \lambda_{y_D}&=\mu b_y\\
\lambda_{z_A}&=\frac{z^*}{J}-\mu b_z, &\qquad \lambda_{z_D}&=\mu b_z\\
\lambda_{x_T}&=-\frac{uc_x}{J}, &\qquad \lambda_{y_T}&=-\frac{uc_y}{J}.
\end{alignedat}
\end{equation}
As $(a_x,a_y,a_z)$, $(b_x,b_y,b_z)$ and $(c_x,c_y)$ are unit vectors, \eqref{eq:lamQ} and \eqref{eq:H0} give
\begin{equation}\label{eq:lamnorm}
\begin{aligned}
\|\bm{\lambda}_D\|_2&=\mu\sqrt{b_x^2+b_y^2+b_z^2}=\mu\\
\|\bm{\lambda}_T\|_2&=\frac{u}{J}\sqrt{c_x^2+c_y^2}=\frac{u}{J}\\
\|\bm{\lambda}_A\|_2&=\|\bm{\lambda}_D\|_2+\gamma\|\bm{\lambda}_T\|_2=\mu+\frac{\gamma u}{J}.
\end{aligned}
\end{equation} As $A$ flies along $(a_x,a_y,a_z)$, \eqref{eq:uopt} reads $\bm{\lambda}_A=-(\mu+\gamma u/J)(a_x,a_y,a_z)$, and equating its components with \eqref{eq:lamQ} gives
\begin{equation}\label{eq:equate}
\begin{aligned}
-\Bigl(\mu+\frac{\gamma u}{J}\Bigr)a_x&=\frac{uc_x}{J}-\mu b_x\\
-\Bigl(\mu+\frac{\gamma u}{J}\Bigr)a_y&=\frac{uc_y}{J}-\mu b_y\\
-\Bigl(\mu+\frac{\gamma u}{J}\Bigr)a_z&=\frac{z^*}{J}-\mu b_z,
\end{aligned}
\end{equation}
which, multiplied by $J$, are \eqref{eq:stat}. Let $g=d_D-r-d_A$, so that $\mathcal{A}^r=\{g=0\}$. By \crefrange{eq:dQ}{eq:JK}, the partial derivatives of $J$ and $g$ at $Q$ are
\begin{equation}\label{eq:gradJ}
\begin{alignedat}{2}
\frac{\partial J}{\partial x}&=\frac{u}{J}\bigl(c_x+\gamma a_x\bigr), &\qquad \frac{\partial g}{\partial x}&=b_x-a_x\\
\frac{\partial J}{\partial y}&=\frac{u}{J}\bigl(c_y+\gamma a_y\bigr), &\qquad \frac{\partial g}{\partial y}&=b_y-a_y\\
\frac{\partial J}{\partial z}&=\frac{\gamma ua_z+z}{J}, &\qquad \frac{\partial g}{\partial z}&=b_z-a_z
\end{alignedat}
\end{equation}
so that \eqref{eq:stat} reads $\nabla J=\mu\nabla g$, the Lagrange condition of $J$ on $\mathcal{A}^r$; among its solutions, $Q^*$ is the one that minimizes $J$.

Next, the value function is verified. As $Q^*$ is a nondegenerate minimizer, it depends smoothly on $\mathbf{x}$ by the implicit function theorem~\cite{rudin1976principles}, and $V$ is $C^1$. The envelope theorem~\cite{milgrom2002envelope}, applied to the Lagrangian $L=J-\mu g$, gives $\partial V/\partial\mathbf{x}=\partial L/\partial\mathbf{x}$ at fixed $Q^*$. The state enters $L$ only through $d_A$, $d_D$ and $d_T$, with
\begin{equation}\label{eq:dpartials}
\begin{alignedat}{2}
\frac{\partial d_A}{\partial\mathbf{x}_A}&=-(a_x,a_y,a_z), &\qquad \frac{\partial d_T}{\partial(x_T,y_T)}&=-(c_x,c_y)\\
\frac{\partial d_D}{\partial\mathbf{x}_D}&=-(b_x,b_y,b_z), &\qquad \frac{\partial J}{\partial(d_A,d_T)}&=\Bigl(\frac{\gamma u}{J},\frac{u}{J}\Bigr),
\end{alignedat}
\end{equation}
and the chain rule gives
\begin{equation}\label{eq:gradV}
\begin{alignedat}{2}
\frac{\partial V}{\partial x_A}&=-\Bigl(\frac{\gamma u}{J}+\mu\Bigr)a_x, &\qquad \frac{\partial V}{\partial x_D}&=\mu b_x\\
\frac{\partial V}{\partial y_A}&=-\Bigl(\frac{\gamma u}{J}+\mu\Bigr)a_y, &\qquad \frac{\partial V}{\partial y_D}&=\mu b_y\\
\frac{\partial V}{\partial z_A}&=-\Bigl(\frac{\gamma u}{J}+\mu\Bigr)a_z, &\qquad \frac{\partial V}{\partial z_D}&=\mu b_z\\
\frac{\partial V}{\partial x_T}&=-\frac{u}{J}c_x, &\qquad \frac{\partial V}{\partial y_T}&=-\frac{u}{J}c_y.
\end{alignedat}
\end{equation}
The Hamiltonian $\partial V/\partial\mathbf{x}\cdot\mathbf{f}$ separates as in \eqref{eq:Hminmax}, and by \eqref{eq:gradV}, with $\mu>0$,
\begin{equation}\label{eq:HJIminmax}
\begin{aligned}
\min_{\beta,\theta_A}\frac{\partial V}{\partial\mathbf{x}_A}\cdot\mathbf{v}_A&=-\Bigl\|\frac{\partial V}{\partial\mathbf{x}_A}\Bigr\|_2=-\frac{\gamma u}{J}-\mu\\
\max_{\delta,\theta_D}\frac{\partial V}{\partial\mathbf{x}_D}\cdot\mathbf{v}_D&=\Bigl\|\frac{\partial V}{\partial\mathbf{x}_D}\Bigr\|_2=\mu\\
\max_{\zeta}\gamma\frac{\partial V}{\partial\mathbf{x}_T}\cdot\mathbf{v}_T&=\gamma\Bigl\|\frac{\partial V}{\partial\mathbf{x}_T}\Bigr\|_2=\frac{\gamma u}{J},
\end{aligned}
\end{equation}
attained by the controls \eqref{eq:strat}: $A$ flies against $\partial V/\partial\mathbf{x}_A$, and $D$ and $T$ along $\partial V/\partial\mathbf{x}_D$ and $\partial V/\partial\mathbf{x}_T$. Substituting \eqref{eq:strat} and \eqref{eq:gradV} into \eqref{eq:dynamics},
\begin{equation}\label{eq:HJIcheck}
\begin{aligned}
\frac{\partial V}{\partial\mathbf{x}}\cdot\mathbf{f}&=-\Bigl(\frac{\gamma u}{J}+\mu\Bigr)\bigl(a_x^2+a_y^2+a_z^2\bigr)\\
&\quad+\mu\bigl(b_x^2+b_y^2+b_z^2\bigr)+\frac{\gamma u}{J}\bigl(c_x^2+c_y^2\bigr)\\
&=-\frac{\gamma u}{J}-\mu+\mu+\frac{\gamma u}{J}=0.
\end{aligned}
\end{equation}
As $V$ does not depend on $t$, \eqref{eq:HJI} holds, and \eqref{eq:strat} is a saddle point.
\end{IEEEproof}

\begin{theorem}[Surface solutions]\label{thm:surface}
For the game \crefrange{eq:dynamics}{eq:value}, the following statements hold.

(i) Under the condition of \cref{thm:air}, let the minimizer $Q^*=(x^*,y^*,0)$ of $J$ lie on $\mathcal{C}^r$, with $(x^*,y^*)\neq(x_T,y_T)$. If there exist $\mu>0$ and $\eta\ge0$ such that
\begin{equation}\label{eq:statS}
\begin{aligned}
c_x+\gamma a_x&=\mu(b_x-a_x),\qquad c_y+\gamma a_y=\mu(b_y-a_y)\\
\gamma a_z&=\mu(b_z-a_z)+\eta,
\end{aligned}
\end{equation}
then the strategies \eqref{eq:strat} with $z^*=0$ are optimal and $V(\mathbf{x})=d_T+\gamma d_A$.

(ii) Under the condition that $\|\mathbf{x}_T-\mathbf{x}_A\|_2<\|\mathbf{x}_T-\mathbf{x}_D\|_2-r$, let $p^*=(x^*,y^*,0)$ be a nondegenerate maximizer of $K$ over $\mathcal{C}^r$ with $K(p^*)>0$, and let $p^*$ minimize $F_{\zeta^*}$ in \eqref{eq:Fz} over $\mathcal{A}^r$, where $\zeta^*$ is the heading from $\mathbf{x}_T$ to $p^*$. If there exist $\mu\ge0$ and $\eta\ge0$ such that
\begin{equation}\label{eq:statC}
\begin{aligned}
\gamma a_x-c_x&=\mu(b_x-a_x),\qquad \gamma a_y-c_y=\mu(b_y-a_y)\\
\gamma a_z&=\mu(b_z-a_z)+\eta,
\end{aligned}
\end{equation}
then $A$ and $D$ follow \eqref{eq:strat} toward $p^*$, $T$ moves toward $p^*$ and through it,
\begin{equation}\label{eq:stratC}
\cos\zeta^*=\frac{x^*-x_T}{d_T},\qquad \sin\zeta^*=\frac{y^*-y_T}{d_T},
\end{equation}
and $V(\mathbf{x})=\gamma d_A-d_T$.

In both cases, $V$ is $C^1$ and satisfies \eqref{eq:HJI}.
\end{theorem}

\begin{IEEEproof}
(i) The proof of \cref{thm:air} applies with $z^*=0$ and $J=u$, except that $\lambda_{z_A}$ in \eqref{eq:lamQ} acquires the term $-\eta$ of \eqref{eq:transA}. Equating $\bm{\lambda}_A=-(\mu+\gamma)(a_x,a_y,a_z)$ with \eqref{eq:lamQ} then gives
\begin{equation}\label{eq:equateS}
\begin{aligned}
-(\mu+\gamma)a_x&=c_x-\mu b_x,\qquad -(\mu+\gamma)a_y=c_y-\mu b_y\\
-(\mu+\gamma)a_z&=-\mu b_z-\eta,
\end{aligned}
\end{equation}
which is \eqref{eq:statS}, the Lagrange condition of $J$ on $\mathcal{A}^r$ under $z\ge0$, with $\eta\ge0$ because moving $Q$ upward along $\mathcal{A}^r$ does not decrease $J$. As $\eta z$ does not depend on $\mathbf{x}$, the gradient \eqref{eq:gradV} holds with $u/J=1$, and \eqref{eq:HJIcheck} again gives zero.

(ii) $T$ starts inside $\mathcal{R}_{AE}\cap\mathcal{S}$ and, by \cref{lem:extremal}, the extremal along which $T$ runs through the point $p\in\mathcal{C}^r$ ends with the payoff $K(p)=\gamma d_A-d_T$, where $K(p)>0$ means that $T$ reaches $p$ before $A$. $T$ chooses the crossing point, which gives the maximizer $p^*$. By \eqref{eq:dQ} and \eqref{eq:cosines}, the partial derivatives of $K$ at $p$ are
\begin{equation}\label{eq:gradK}
\begin{aligned}
\frac{\partial K}{\partial x}&=\gamma a_x-c_x,\qquad \frac{\partial K}{\partial y}=\gamma a_y-c_y\\
\frac{\partial K}{\partial z}&=\gamma a_z,
\end{aligned}
\end{equation}
so that, with \eqref{eq:gradJ}, the first two equations of \eqref{eq:statC} are the Lagrange condition of $K$ on $\mathcal{C}^r$. $A$ replies with $p^*$ by assumption; the third equation, with $\eta\ge0$, states that lifting the interception point along $\mathcal{A}^r$ does not lower the payoff. The envelope theorem, applied to $K-\mu g$ with $\partial K/\partial d_A=\gamma$ and $\partial K/\partial d_T=-1$, now gives
\begin{equation}\label{eq:gradVC}
\begin{alignedat}{2}
\frac{\partial V}{\partial x_A}&=-(\gamma+\mu)a_x, &\qquad \frac{\partial V}{\partial x_D}&=\mu b_x\\
\frac{\partial V}{\partial y_A}&=-(\gamma+\mu)a_y, &\qquad \frac{\partial V}{\partial y_D}&=\mu b_y\\
\frac{\partial V}{\partial z_A}&=-(\gamma+\mu)a_z, &\qquad \frac{\partial V}{\partial z_D}&=\mu b_z\\
\frac{\partial V}{\partial x_T}&=c_x, &\qquad \frac{\partial V}{\partial y_T}&=c_y.
\end{alignedat}
\end{equation}
Since $\mu\ge0$, the strategies of $A$, $D$ and \eqref{eq:stratC} attain the minimum and the maxima in \eqref{eq:HJI}, and
\begin{equation}\label{eq:HJIcheckC}
\begin{aligned}
\frac{\partial V}{\partial\mathbf{x}}\cdot\mathbf{f}&=-(\gamma+\mu)\bigl(a_x^2+a_y^2+a_z^2\bigr)+\mu\bigl(b_x^2+b_y^2+b_z^2\bigr)\\
&\quad+\gamma\bigl(c_x^2+c_y^2\bigr)=-\gamma-\mu+\mu+\gamma=0,
\end{aligned}
\end{equation}
which completes the proof.
\end{IEEEproof}

\begin{remark}\label{rem:planar}
If $T$ lies on $\mathcal{C}^r$ and \cref{thm:air,thm:surface} apply on both sides, the two payoffs meet at $p=Q=\mathbf{x}_T$, and $V(\mathbf{x})=\gamma\|\mathbf{x}_T-\mathbf{x}_A\|_2$ from both sides. If $z_{A_0}=z_{D_0}=0$ and $\theta_A=\theta_D\equiv0$, $\mathcal{C}^r$ becomes the interception hyperbola, and \cref{thm:surface}(i), \cref{thm:surface}(ii) and the boundary case reduce to Theorems~1, 2 and 3 of~\cite{shi2026constructing}. \Cref{thm:air} and \cref{thm:aircross} below have no planar counterpart. Moreover, $\partial V/\partial x_D$, $\partial V/\partial y_D$ and $\partial V/\partial z_D$ do not vanish, because the multiplier $\mu=\partial V/\partial r$ of the interception constraint is positive.
\end{remark}

\begin{corollary}\label{cor:plane}
In \cref{thm:air}, $Q^*$ does not lie in the plane $\Pi$ through $\mathbf{x}_A$, $\mathbf{x}_D$ and $\mathbf{x}_T$ unless $\Pi$ is vertical.
\end{corollary}

\begin{IEEEproof}
Suppose $Q^*\in\Pi$, and let $(n_x,n_y,n_z)$ be a unit normal of $\Pi$. As $\mathcal{A}^r$ is a surface of revolution about the line through $\mathbf{x}_A$ and $\mathbf{x}_D$, which lies in $\Pi$, the normal of $\Pi$ is tangent to $\mathcal{A}^r$ at $Q^*$, and \eqref{eq:stat} requires $\partial J^2/\partial n=0$. Since $Q^*$, $\mathbf{x}_A$ and $\mathbf{x}_T$ lie in $\Pi$, $a_xn_x+a_yn_y+a_zn_z=0$ and $d_T(c_xn_x+c_yn_y)=-z^*n_z$, so that, by \eqref{eq:gradJ},
\begin{equation}\label{eq:offplane}
\begin{aligned}
\frac{\partial J^2}{\partial n}&=2u\bigl(c_xn_x+c_yn_y\bigr)+2z^*n_z\\
&\quad+2\gamma u\bigl(a_xn_x+a_yn_y+a_zn_z\bigr)\\
&=-\frac{2uz^*n_z}{d_T}+2z^*n_z=-\frac{2\gamma d_Az^*n_z}{d_T}\neq0
\end{aligned}
\end{equation}
whenever $n_z\neq0$, because $\gamma>0$ and $z^*>0$. This contradiction proves the claim.
\end{IEEEproof}

\begin{figure}[!t]
\centering
\includegraphics[width=\columnwidth]{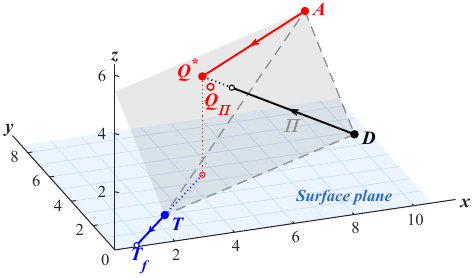}
\caption{Evasion in the air of \cref{ex:straight} for $\mathbf{x}_A=(8.4,5.6,4.6)$, $\mathbf{x}_D=(8.6,1.6,2)$, $\mathbf{x}_T=(2.4,2,0)$, $\gamma=0.6$ and $r=1$. $A$ and $D$ fly straight to $Q^*$, and $D$ stops at the distance $r$ from it. $T$ flees from the foot of $Q^*$ to $T_f$. The grey parallelogram is a part of the plane $\Pi$ through $A$, $D$ and $T$ (dashed triangle); the minimizer $Q_\Pi$ of $J$ restricted to $\Pi$ differs from $Q^*$, which lies off $\Pi$.}
\label{fig:air}
\end{figure}

\begin{example}\label{ex:straight}
Let $\mathbf{x}_A=(8.4,5.6,4.6)$ and $\mathbf{x}_D=(8.6,1.6,2)$, as in \cref{fig:scenario}, with $\gamma=0.6$ and $r=1$. For $\mathbf{x}_T=(2.4,2,0)$, \cref{thm:air} gives $Q^*=(4.5698,4.5128,3.3866)$, $\mu=0.6283$ and $V^*=6.7313$, reached at $t_f-t_0=4.1623$ (\cref{fig:air}). The optimal headings are $\beta^*=-164.15^\circ$, $\theta_A^*=-16.95^\circ$, $\delta^*=144.14^\circ$, $\theta_D^*=15.58^\circ$ and $\zeta^*=-130.81^\circ$. The point $Q^*$ lies $0.7270$ from the plane $\Pi$; restricting the interception point to $\Pi$, as the planar reduction would do, raises the payoff to $6.8044$. With $\mathbf{x}_A=(9,1,1)$ and $\mathbf{x}_D=(5,3,6)$, the defender above the attacker drives the interception point to the surface. For $\mathbf{x}_T=(0,0,0)$, \cref{thm:surface}(i) gives $Q^*=(3.1874,-0.1275,0)$, $\mu=0.5899$, $\eta=0.3072$ and $V^*=6.7928$; for $\mathbf{x}_T=(12,13,0)$, inside $\mathcal{R}_{AE}\cap\mathcal{S}$, \cref{thm:surface}(ii) gives $p^*=(11.7893,13.6045,0)$, $\mu=1.6896$, $\eta=0.5500$ and $V^*=7.1287$, and no point of $\mathcal{A}^r$ in the air is a better reply of $A$.
\end{example}

\subsection{Crossing in the Air}\label{subsec:overhead}

\Cref{thm:air,thm:surface} exclude interception points directly above $T$, where $d_T=0$ and $J$ has a conical kink. Such points are not exceptional, and they lead to a third family of solutions.

\begin{definition}\label{def:overhead}
Let $\|\mathbf{x}_T-\mathbf{x}_A\|_2>\|\mathbf{x}_T-\mathbf{x}_D\|_2-r$. The overhead point $Q_c=(x_T,y_T,z_c)$ is the lowest point of $\mathcal{A}^r$ on the vertical line through $\mathbf{x}_T$.
\end{definition}

At $Q_c$, $J(Q_c)=\sqrt{\gamma^2d_A^2+z_c^2}$. Let $\mu_c$, $\lambda_x$ and $\lambda_y$ solve
\begin{equation}\label{eq:statO}
\begin{aligned}
\frac{\gamma^2d_A}{J}a_x&=\mu_c(b_x-a_x)+\lambda_x\\
\frac{\gamma^2d_A}{J}a_y&=\mu_c(b_y-a_y)+\lambda_y\\
\frac{\gamma^2d_A}{J}a_z+\frac{z_c}{J}&=\mu_c(b_z-a_z)
\end{aligned}
\end{equation}
where \eqref{eq:dQ} and \eqref{eq:cosines} are evaluated at $Q_c$, $J=J(Q_c)$, and $b_z\neq a_z$. Solving the third equation of \eqref{eq:statO} for $\mu_c$ and substituting it into the first two give
\begin{equation}\label{eq:statOsol}
\begin{aligned}
\mu_c&=\frac{\gamma^2d_Aa_z+z_c}{J(b_z-a_z)}\\
\lambda_x&=\frac{\gamma^2d_A}{J}a_x-\mu_c(b_x-a_x)\\
\lambda_y&=\frac{\gamma^2d_A}{J}a_y-\mu_c(b_y-a_y).
\end{aligned}
\end{equation}

\begin{proposition}\label{prop:overhead}
$Q_c$ is a strict local minimizer of $J$ on $\mathcal{A}^r$ if
\begin{equation}\label{eq:band}
\sqrt{\lambda_x^2+\lambda_y^2}<\frac{\gamma d_A}{J},
\end{equation}
and it is not a local minimizer if the inequality is reversed. The set of states satisfying \eqref{eq:band} is open.
\end{proposition}

\begin{IEEEproof}
Let $\bm{\tau}=(\tau_x,\tau_y,\tau_z)$ be tangent to $\mathcal{A}^r$ at $Q_c$, so that $(b_x-a_x)\tau_x+(b_y-a_y)\tau_y+(b_z-a_z)\tau_z=0$. As $b_z\neq a_z$,
\begin{equation}\label{eq:tauz}
\tau_z=-\frac{(b_x-a_x)\tau_x+(b_y-a_y)\tau_y}{b_z-a_z}
\end{equation}
is fixed by $(\tau_x,\tau_y)$, which ranges over $\mathbb{R}^2$. Moving $Q$ from $Q_c$ along $\bm{\tau}$ changes $d_T$ at the rate $\sqrt{\tau_x^2+\tau_y^2}$, so that, by \eqref{eq:JK} and \eqref{eq:statO}, the one-sided derivative of $J$ is
\begin{equation}\label{eq:dJtau}
\medmuskip=7mu\relax \thickmuskip=8mu\relax
\begin{aligned}
J'(Q_c;\bm{\tau})\enspace&=\frac{\gamma d_A}{J}\,\sqrt{\tau_x^2+\tau_y^2}\:+\:\frac{z_c}{J}\,\tau_z\\
&\quad+\frac{\gamma^2d_A}{J}\,\bigl(a_x\,\tau_x+a_y\,\tau_y+a_z\,\tau_z\bigr)\\
&=\frac{\gamma d_A}{J}\,\sqrt{\tau_x^2+\tau_y^2}\:+\:\lambda_x\,\tau_x\:+\:\lambda_y\,\tau_y.
\end{aligned}
\end{equation}
Its minimum over the unit circle of $(\tau_x,\tau_y)$ is $\gamma d_A/J-\sqrt{\lambda_x^2+\lambda_y^2}$. If \eqref{eq:band} holds, $J'(Q_c;\bm{\tau})>0$ for every $\bm{\tau}\neq\mathbf{0}$, and $Q_c$ is a strict local minimizer; if the inequality is reversed, $J$ decreases along some $\bm{\tau}$. The solution of the linear system \eqref{eq:statO} depends continuously on $\mathbf{x}$, so \eqref{eq:band} defines an open set.
\end{IEEEproof}

\begin{proposition}\label{thm:overhead}
For the game \crefrange{eq:dynamics}{eq:value}, let \eqref{eq:band} hold and let $Q_c$ minimize $J$ on $\mathcal{A}^r$. Then $V^+(\mathbf{x})=J(Q_c)$ satisfies
\begin{equation}\label{eq:deficit}
\min_{\beta,\theta_A}\,\max_{\delta,\theta_D,\zeta}\frac{\partial V^+}{\partial\mathbf{x}}\cdot\mathbf{f}=\gamma\Bigl(\sqrt{\lambda_x^2+\lambda_y^2}-\frac{\gamma d_A}{J}\Bigr)<0.
\end{equation}
Hence $V^+$ violates the HJI equation, and the straight motions toward $Q_c$ are not a saddle point.
\end{proposition}

\begin{IEEEproof}
On the overhead region, $Q_c$ is the minimizer of $J$ under the additional constraint $(x,y)=(x_T,y_T)$, whose multipliers are $\lambda_x$ and $\lambda_y$ in \eqref{eq:statO}. The Lagrangian
\begin{equation}\label{eq:Lc}
L_c=J-\mu_c\,g-\lambda_x(x-x_T)-\lambda_y(y-y_T),
\end{equation}
with $d_T=0$ and $\partial J/\partial d_A=\gamma^2d_A/J$ at $Q_c$, has $\partial L_c/\partial x_T=\lambda_x$ and $\partial L_c/\partial y_T=\lambda_y$, and the envelope theorem gives
\begin{equation}\label{eq:gradVO}
\begin{alignedat}{2}
\frac{\partial V^+}{\partial x_A}&=-\Bigl(\frac{\gamma^2d_A}{J}+\mu_c\Bigr)a_x, &\qquad \frac{\partial V^+}{\partial x_D}&=\mu_cb_x\\
\frac{\partial V^+}{\partial y_A}&=-\Bigl(\frac{\gamma^2d_A}{J}+\mu_c\Bigr)a_y, &\qquad \frac{\partial V^+}{\partial y_D}&=\mu_cb_y\\
\frac{\partial V^+}{\partial z_A}&=-\Bigl(\frac{\gamma^2d_A}{J}+\mu_c\Bigr)a_z, &\qquad \frac{\partial V^+}{\partial z_D}&=\mu_cb_z\\
\frac{\partial V^+}{\partial x_T}&=\lambda_x, &\qquad \frac{\partial V^+}{\partial y_T}&=\lambda_y.
\end{alignedat}
\end{equation}
With $\mu_c>0$, the minimum over the controls of $A$ and the maxima over those of $D$ and $T$ give
\begin{equation}\label{eq:deficit1}
\medmuskip=0.8mu\relax \thickmuskip=2.5mu\relax
-\bigl(\tfrac{\gamma^2d_A}{J}+\mu_c\bigr)+\mu_c+\gamma\sqrt{\lambda_x^2+\lambda_y^2}
=\gamma\bigl(\sqrt{\lambda_x^2+\lambda_y^2}-\tfrac{\gamma d_A}{J}\bigr),
\end{equation}
which is negative by \eqref{eq:band}. Thus, whatever $D$ and $T$ do, $A$ can make $V^+$ decrease, and the straight motions toward $Q_c$ do not form a saddle point.
\end{IEEEproof}

At $Q_c$, the evasion payoff $J$ and the crossing payoff $\sqrt{K^2+z^2}$ of \cref{lem:extremal} coincide, since $d_T=0$. Committed to $Q_c$, $A$ leaves $T$ free to flee in any direction. Waiting for the heading of $T$, $A$ aims instead at a point slightly ahead of $T$, under which $T$ then runs. This crossing in the air also replaces the crossing on the surface when $\eta<0$ in \eqref{eq:statC}, that is, when $A$ gains by lifting the interception point.

\begin{theorem}[Crossing in the air]\label{thm:aircross}
For the game \crefrange{eq:dynamics}{eq:value}, let $\zeta^*$ and $Q^*=(x^*,y^*,z^*)$ solve the max--min in \eqref{eq:Vpm} with $z^*>0$ and $(x^*,y^*)=(x_T,y_T)+d_T(\cos\zeta^*,\sin\zeta^*)$, where $0<d_T<\gamma d_A$, so that $T$ passes under $Q^*$ before $t_f$. If the solution is nondegenerate and the multiplier $\mu$ in
\begin{equation}\label{eq:statX}
\begin{aligned}
K(\gamma a_x-c_x)&=\mu P(b_x-a_x)\\
K(\gamma a_y-c_y)&=\mu P(b_y-a_y)\\
K\gamma a_z+z^*&=\mu P(b_z-a_z)
\end{aligned}
\end{equation}
is positive, where $K=\gamma d_A-d_T$ and $P=\sqrt{K^2+z^{*2}}$, then $A$ and $D$ follow \eqref{eq:strat} toward $Q^*$, $T$ moves as in \eqref{eq:stratC} and passes under $Q^*$, and
\begin{equation}\label{eq:Vcross}
V(\mathbf{x})=V^-(\mathbf{x})=\sqrt{(\gamma d_A-d_T)^2+z^{*2}}.
\end{equation}
$V$ is $C^1$ and satisfies \eqref{eq:HJI}.
\end{theorem}

\begin{IEEEproof}
By \cref{lem:extremal}, the extremal along which $T$ passes under $Q^*$ ends with the payoff $P=F_{\zeta^*}(Q^*)$. The stationarity of \eqref{eq:Vpm} in $\zeta$ reads
\begin{equation}\label{eq:dFzeta}
\frac{\partial F_\zeta^2}{\partial\zeta}=2\gamma d_A\bigl[(x-x_T)\sin\zeta-(y-y_T)\cos\zeta\bigr]=0,
\end{equation}
so the foot of $Q^*$ lies on the path of $T$, at the distance $d_T$ ahead of $\mathbf{x}_T$. At $Q^*$, $F_{\zeta^*}=P$, and by \eqref{eq:dFzeta} their gradients coincide. The partial derivatives of $P$ at $Q$ are
\begin{equation}\label{eq:gradP}
\begin{aligned}
\frac{\partial P}{\partial x}&=\frac{K}{P}\bigl(\gamma a_x-c_x\bigr),\qquad \frac{\partial P}{\partial y}=\frac{K}{P}\bigl(\gamma a_y-c_y\bigr)\\
\frac{\partial P}{\partial z}&=\frac{K\gamma a_z+z}{P},
\end{aligned}
\end{equation}
so that the stationarity of \eqref{eq:Vpm} in $Q$, $\nabla P=\mu\nabla g$ with \eqref{eq:gradJ}, is \eqref{eq:statX}. The maximum principle leads to the same conditions. Along this extremal, $x_{A_f}-x_{T_f}=-Kc_x$, $y_{A_f}-y_{T_f}=-Kc_y$ and $z_{A_f}=z^*$, and \crefrange{eq:transA}{eq:transT} give, with $\mu=-\nu$ and $\Phi_f=P$,
\begin{equation}\label{eq:lamX}
\begin{alignedat}{2}
\lambda_{x_A}&=-\frac{Kc_x}{P}-\mu b_x, &\qquad \lambda_{x_D}&=\mu b_x\\
\lambda_{y_A}&=-\frac{Kc_y}{P}-\mu b_y, &\qquad \lambda_{y_D}&=\mu b_y\\
\lambda_{z_A}&=\frac{z^*}{P}-\mu b_z, &\qquad \lambda_{z_D}&=\mu b_z\\
\lambda_{x_T}&=\frac{Kc_x}{P}, &\qquad \lambda_{y_T}&=\frac{Kc_y}{P}.
\end{alignedat}
\end{equation}
Hence $\|\bm{\lambda}_D\|_2=\mu$, $\|\bm{\lambda}_T\|_2=K/P$, and \eqref{eq:H0} gives $\|\bm{\lambda}_A\|_2=\mu+\gamma K/P$; equating $\bm{\lambda}_A=-(\mu+\gamma K/P)(a_x,a_y,a_z)$ with \eqref{eq:lamX} and multiplying by $P$ again yields \eqref{eq:statX}. As the solution is nondegenerate, the envelope theorem~\cite{milgrom2002envelope}, applied to $P-\mu g$ with $\partial P/\partial d_A=\gamma K/P$ and $\partial P/\partial d_T=-K/P$, gives
\begin{equation}\label{eq:gradVX}
\begin{alignedat}{2}
\frac{\partial V}{\partial x_A}&=-\Bigl(\frac{\gamma K}{P}+\mu\Bigr)a_x, &\qquad \frac{\partial V}{\partial x_D}&=\mu b_x\\
\frac{\partial V}{\partial y_A}&=-\Bigl(\frac{\gamma K}{P}+\mu\Bigr)a_y, &\qquad \frac{\partial V}{\partial y_D}&=\mu b_y\\
\frac{\partial V}{\partial z_A}&=-\Bigl(\frac{\gamma K}{P}+\mu\Bigr)a_z, &\qquad \frac{\partial V}{\partial z_D}&=\mu b_z\\
\frac{\partial V}{\partial x_T}&=\frac{K}{P}c_x, &\qquad \frac{\partial V}{\partial y_T}&=\frac{K}{P}c_y.
\end{alignedat}
\end{equation}
Since $\mu>0$, as in \eqref{eq:HJIminmax},
\begin{equation}\label{eq:HJIminmaxX}
\begin{aligned}
\min_{\beta,\theta_A}\frac{\partial V}{\partial\mathbf{x}_A}\cdot\mathbf{v}_A&=-\Bigl\|\frac{\partial V}{\partial\mathbf{x}_A}\Bigr\|_2=-\frac{\gamma K}{P}-\mu\\
\max_{\delta,\theta_D}\frac{\partial V}{\partial\mathbf{x}_D}\cdot\mathbf{v}_D&=\Bigl\|\frac{\partial V}{\partial\mathbf{x}_D}\Bigr\|_2=\mu\\
\max_{\zeta}\gamma\frac{\partial V}{\partial\mathbf{x}_T}\cdot\mathbf{v}_T&=\gamma\Bigl\|\frac{\partial V}{\partial\mathbf{x}_T}\Bigr\|_2=\frac{\gamma K}{P},
\end{aligned}
\end{equation}
attained when $A$ and $D$ fly toward $Q^*$ and $T$ runs toward its foot, and
\begin{equation}\label{eq:HJIcheckX}
\begin{aligned}
\frac{\partial V}{\partial\mathbf{x}}\cdot\mathbf{f}&=-\Bigl(\frac{\gamma K}{P}+\mu\Bigr)\bigl(a_x^2+a_y^2+a_z^2\bigr)\\
&\quad+\mu\bigl(b_x^2+b_y^2+b_z^2\bigr)+\frac{\gamma K}{P}\bigl(c_x^2+c_y^2\bigr)\\
&=-\frac{\gamma K}{P}-\mu+\mu+\frac{\gamma K}{P}=0.
\end{aligned}
\end{equation}
Finally, $T$ guarantees $V^-$ by \eqref{eq:Vpm}, and along the strategy of $A$, $V$ does not increase by \eqref{eq:HJIcheckX} while $\mathbf{x}$ stays in the region of \crefrange{thm:air}{thm:aircross}. Hence $V=V^-$.
\end{IEEEproof}

\begin{remark}\label{rem:overhead}
\Crefrange{thm:air}{thm:aircross} give four regimes: the interception point lies in the air or on the surface, and $T$ either flees from its foot or runs under it. The overhead point marks the junction of the evasion and the crossing in the air, where $d_T=0$ and $J=P$; on the band of \cref{prop:overhead}, the crossing has $d_T>0$ and lies strictly below $V^+$. In the plane, the interception point coincides with $T$ only when $T$ lies on the interception hyperbola~\cite[Thm.~3]{shi2026constructing}, and $T$ crosses the hyperbola only from inside it~\cite[Thm.~2]{shi2026constructing}; a pursuer directly above a planar evader is likewise a degenerate case in~\cite{li2022different}. In space, $T$ runs under the attacker from both sides of $\mathcal{C}^r$.
\end{remark}

\begin{figure}[!t]
\centering
\includegraphics[width=\columnwidth]{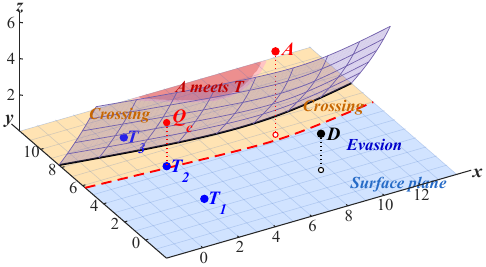}
\caption{Solution regimes of the game of degree over the position of $T$ for $\mathbf{x}_A=(8.4,5.6,4.6)$, $\mathbf{x}_D=(8.6,1.6,2)$, $\gamma=0.6$ and $r=1$. The interception surface $\mathcal{A}^r$ (translucent) meets $\mathcal{S}$ in the trace $\mathcal{C}^r$. $T$ runs under a point of $\mathcal{A}^r$ in the air in the orange region (\cref{thm:aircross}), through a point of $\mathcal{C}^r$ in the red region (\cref{thm:surface}(ii)), and away from the foot of the interception point in the blue region (\cref{thm:air}); in the dark red region, $A$ meets $T$. The dashed curve bounds the overhead band of \cref{prop:overhead}. $T_1$, $T_2$ and $T_3$ are the targets of \cref{ex:straight,ex:regimes}.}
\label{fig:regimes}
\end{figure}

\begin{figure}[!t]
\centering
\includegraphics[width=\columnwidth]{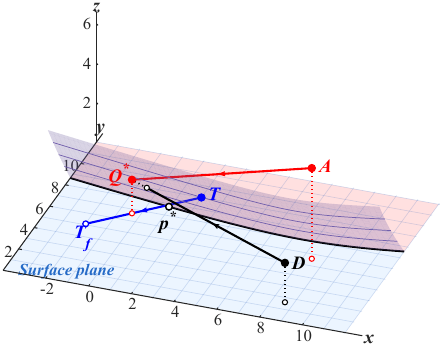}
\caption{Crossing in the air of \cref{ex:regimes} at $T_3=(2,9,0)$, inside $\mathcal{C}^r$, for the attacker and defender of \cref{fig:regimes}. $A$ and $D$ fly straight to $Q^*$, and $D$ stops at the distance $r$ from it. $T$ runs out of $\mathcal{C}^r$, passes under $Q^*$ and reaches $T_f$. The hollow point $p^*$ on $\mathcal{C}^r$ is the crossing point on the surface, which $A$ does not choose since $\eta<0$.}
\label{fig:aircross}
\end{figure}

\begin{example}\label{ex:regimes}
\Cref{fig:regimes} classifies the target positions for the attacker and defender of \cref{ex:straight}. In its window, the crossing in the air of \cref{thm:aircross} covers $42.2\%$ of the plane, on the overhead band of \cref{prop:overhead} and on most of $\mathcal{R}_{AE}\cap\mathcal{S}$; the evasion in the air of \cref{thm:air} covers $53.1\%$, the crossing on the surface of \cref{thm:surface}(ii) only $0.8\%$, and on the remaining $3.9\%$, near the foot of $A$, $A$ meets $T$ wherever it runs. At $T_2=(2.4,5.6,0)$, the state of \cref{fig:scenario}, $Q_c=(2.4,5.6,2.4036)$, $V^+=4.5248$, $\mu_c=0.8948$ and $\sqrt{\lambda_x^2+\lambda_y^2}=0.7456<\gamma d_A/J=0.8472$, so that \eqref{eq:deficit} equals $-0.0610$. \Cref{thm:aircross} gives $\zeta^*=-134.49^\circ$ and $Q^*=(2.2903,5.4883,2.5843)$: $T$ passes under $Q^*$ after $d_T=0.1565$, and $V=4.5166$ with $\mu=1.0668$. At $T_3=(2,9,0)$, inside $\mathcal{C}^r$, the maximizer of $K$ over $\mathcal{C}^r$ is $p^*=(0.9368,7.7485,0)$ with $K=3.7736$, but $\eta=-1.0539$ in \eqref{eq:statC}. \Cref{thm:aircross} gives $\zeta^*=-135.50^\circ$, $Q^*=(-0.3935,6.6479,1.7028)$, $d_T=3.3558$, $\mu=1.6221$ and $V=2.8096$ (\cref{fig:aircross}). At both states, the central differences of $V$ satisfy \eqref{eq:HJI} to $10^{-10}$. In closed loop, every tested deviation of $T$ lowers the payoff below $V$, e.g., to $2.8055$ when $T$ runs through $p^*$, and every tested deviation of $A$ raises it, e.g., to $3.9573$ when $A$ flies to $p^*$.
\end{example}

\Crefrange{thm:air}{thm:aircross} give the value $V$ wherever the interception point is regular. \Cref{sec:kind} uses its level set $V=R$ as the barrier of the game of kind.

\section{Game of Kind}\label{sec:kind}

\Cref{prob:kind} asks whether $A$ can drive $\mathbf{x}$ into $\mathcal{M}_1$ before $\mathcal{M}_2$. For $R=0$, the barrier is found by comparing the Apollonius circle of $T$--$A$ with the interception curve of $D$--$A$~\cite{garcia2018cooperative,garcia2019design,liang2019differential}. For $R>0$, the limit interception curve of~\cite{shi2025determination,shi2026constructing} extends the interception curve by $R$ along straight strikes of $A$. This section obtains the barrier from the value function of \cref{sec:degree}, shows that the straight-strike construction is not semipermeable for $R>0$, and then studies the winning regions on the target plane.

\subsection{Barrier}\label{subsec:barrier}

Along the optimal motions of \crefrange{thm:air}{thm:aircross}, the payoff is the $A$--$T$ distance at interception. The following lemma tells when this distance is also the smallest distance of the play.

\begin{lemma}\label{lem:monotone}
Along the optimal motions of \crefrange{thm:air}{thm:aircross}, $\|\mathbf{x}_A-\mathbf{x}_T\|_2$ is a convex function of $t$, and it decreases on $[t_0,t_f]$ if
\begin{equation}\label{eq:monotone}
\begin{aligned}
(d_T+\gamma d_A)(c_xa_x+c_ya_y+\gamma)+z^*a_z&<0\\
K(c_xa_x+c_ya_y-\gamma)-z^*a_z&>0
\end{aligned}
\end{equation}
for the evasion and the crossing solutions, respectively, where $K=\gamma d_A-d_T$ and \eqref{eq:dQ} and \eqref{eq:cosines} are evaluated at the interception point. In that case, $\|\mathbf{x}_A-\mathbf{x}_T\|_2\ge V(\mathbf{x})$.
\end{lemma}

\begin{IEEEproof}
All players move straight, so $\mathbf{x}_A(t)-\mathbf{x}_T(t)$ is affine in $t$, and its norm $\sigma(t)=\|\mathbf{x}_A(t)-\mathbf{x}_T(t)\|_2$ is convex, with
\begin{equation}\label{eq:dist0}
\dot{\sigma}=\frac{(\mathbf{x}_A-\mathbf{x}_T)\cdot(\dot{\mathbf{x}}_A-\dot{\mathbf{x}}_T)}{\|\mathbf{x}_A-\mathbf{x}_T\|_2}.
\end{equation}
A convex function that decreases at $t_f$ decreases on $[t_0,t_f]$. For the evasion solutions, the terminal differences are \eqref{eq:diffs}, with $u=d_T+\gamma d_A$; the velocity of $A$ is $(a_x,a_y,a_z)$ and that of $T$ is $-\gamma(c_x,c_y)$. Hence, as $c_x^2+c_y^2=1$,
\begin{equation}\label{eq:dist1}
\begin{aligned}
\dot{\sigma}(t_f)&=\frac{uc_x(a_x+\gamma c_x)+uc_y(a_y+\gamma c_y)+z^*a_z}{J}\\
&=\frac{u(c_xa_x+c_ya_y+\gamma)+z^*a_z}{J}.
\end{aligned}
\end{equation}
For the crossing solutions, $x_{A_f}-x_{T_f}=-Kc_x$, $y_{A_f}-y_{T_f}=-Kc_y$ and $z_{A_f}=z^*$, and $T$ moves along $\gamma(c_x,c_y)$, so that, with $P=\sqrt{K^2+z^{*2}}$,
\begin{equation}\label{eq:dist2}
\begin{aligned}
\dot{\sigma}(t_f)&=\frac{-Kc_x(a_x-\gamma c_x)-Kc_y(a_y-\gamma c_y)+z^*a_z}{P}\\
&=\frac{\gamma K+z^*a_z-K(c_xa_x+c_ya_y)}{P}.
\end{aligned}
\end{equation}
Under \eqref{eq:monotone}, both derivatives are negative; the distance therefore exceeds its terminal value $V(\mathbf{x})$ at every earlier time.
\end{IEEEproof}

Let $\Omega$ denote the open set of states at which one of \crefrange{thm:air}{thm:aircross} applies and \eqref{eq:monotone} holds.

\begin{theorem}[Barrier]\label{thm:barrier}
For the game \crefrange{eq:dynamics}{eq:M}, let $\mathbf{x}_0\in\Omega$.
(i) If $V(\mathbf{x}_0)>R$, the strategies of $D$ and $T$ in \crefrange{thm:air}{thm:aircross} keep $\|\mathbf{x}_A-\mathbf{x}_T\|_2>R$ as long as $\mathbf{x}$ stays in $\Omega$, whatever $A$ does.
(ii) If $V(\mathbf{x}_0)<R$, the strategy of $A$ in \crefrange{thm:air}{thm:aircross} captures $T$ no later than $D$ intercepts $A$, as long as $\mathbf{x}$ stays in $\Omega$, whatever $D$ and $T$ do.
(iii) Hence the barrier on $\Omega$ is
\begin{equation}\label{eq:barrier}
\mathcal{B}\cap\Omega=\bigl\{\mathbf{x}\in\Omega\bigm|V(\mathbf{x})=R\bigr\},
\end{equation}
and it is semipermeable: $\min_{\beta,\theta_A}\max_{\delta,\theta_D,\zeta}\partial V/\partial\mathbf{x}\cdot\mathbf{f}=0$ on $\mathcal{B}\cap\Omega$.
\end{theorem}

\begin{IEEEproof}
The Hamiltonian $\partial V/\partial\mathbf{x}\cdot\mathbf{f}$ is a sum of a term in the controls of $A$, a term in those of $D$ and a term in that of $T$, as in \eqref{eq:Hminmax}. By \eqref{eq:gradV}, \eqref{eq:gradVC} and \eqref{eq:gradVX}, when $D$ and $T$ use their strategies,
\begin{equation}\label{eq:Vdot1}
\begin{aligned}
\frac{\mathrm{d}V}{\mathrm{d}t}&=\frac{\partial V}{\partial\mathbf{x}_A}\cdot\mathbf{v}_A+\Bigl\|\frac{\partial V}{\partial\mathbf{x}_D}\Bigr\|_2+\gamma\Bigl\|\frac{\partial V}{\partial\mathbf{x}_T}\Bigr\|_2\\
&\ge-\Bigl\|\frac{\partial V}{\partial\mathbf{x}_A}\Bigr\|_2+\Bigl\|\frac{\partial V}{\partial\mathbf{x}_D}\Bigr\|_2+\gamma\Bigl\|\frac{\partial V}{\partial\mathbf{x}_T}\Bigr\|_2=0
\end{aligned}
\end{equation}
for every control of $A$, and when $A$ uses its strategy,
\begin{equation}\label{eq:Vdot2}
\begin{aligned}
\frac{\mathrm{d}V}{\mathrm{d}t}&=-\Bigl\|\frac{\partial V}{\partial\mathbf{x}_A}\Bigr\|_2+\frac{\partial V}{\partial\mathbf{x}_D}\cdot\mathbf{v}_D+\gamma\frac{\partial V}{\partial\mathbf{x}_T}\cdot\mathbf{v}_T\\
&\le-\Bigl\|\frac{\partial V}{\partial\mathbf{x}_A}\Bigr\|_2+\Bigl\|\frac{\partial V}{\partial\mathbf{x}_D}\Bigr\|_2+\gamma\Bigl\|\frac{\partial V}{\partial\mathbf{x}_T}\Bigr\|_2=0
\end{aligned}
\end{equation}
for every control of $D$ and $T$, where the last equalities are the HJI equation \eqref{eq:HJI} in the form \eqref{eq:HJIminmax}. (i) By \eqref{eq:Vdot1} and \cref{lem:monotone}, $\|\mathbf{x}_A-\mathbf{x}_T\|_2\ge V(\mathbf{x}(t))\ge V(\mathbf{x}_0)>R$ while $\mathbf{x}\in\Omega$, so $\mathcal{M}_1$ is not reached. (ii) By \eqref{eq:Vdot2}, $V(\mathbf{x}(t))\le V(\mathbf{x}_0)<R$. When $\mathbf{x}$ reaches $\mathcal{M}_2$, the play is over, and $V$ equals the $A$--$T$ distance by \eqref{eq:payoff}; hence this distance is below $R$, and $\mathcal{M}_1$ has been reached first. If $D$ never intercepts $A$, $A$ is unimpeded and captures $T$ in finite time, as $\gamma<1$. (iii) follows from (i) and (ii), and the semipermeability~\cite{isaacs1965differential,basar1998dynamic} follows from \eqref{eq:HJI}.
\end{IEEEproof}

For fixed $\mathbf{x}_A$ and $\mathbf{x}_D$, the barrier is a curve on the target plane. Where the evasion or the crossing on the surface applies, it is the envelope of two families of disks on $\mathcal{S}$: the capture disks of $A$, centered at the feet of the points $Q=(x_Q,y_Q,z_Q)$ of $\mathcal{A}^r$ with $z_Q<R$, and the escape disks of $T$, centered at the points $p=(x_p,y_p,0)$ of $\mathcal{C}^r$,
\begin{equation}\label{eq:disks}
\begin{aligned}
\mathcal{D}_Q&=\bigl\{\mathbf{x}_T\in\mathcal{S}\bigm|(x_T-x_Q)^2+(y_T-y_Q)^2<\varrho_Q^2\bigr\}\\
\mathcal{E}_p&=\bigl\{\mathbf{x}_T\in\mathcal{S}\bigm|(x_T-x_p)^2+(y_T-y_p)^2<\varrho_p^2\bigr\}\\
\varrho_Q&=\sqrt{R^2-z_Q^2}-\gamma d_A,\qquad \varrho_p=\gamma d_A-R
\end{aligned}
\end{equation}
where $d_A$ is the distance from $\mathbf{x}_A$ to $Q$ or to $p$, and only positive radii are kept.

\begin{corollary}\label{cor:disks}
Let $\mathbf{x}\in\Omega$. Then $A$ wins if and only if $V^-(\mathbf{x})<R$ in \eqref{eq:Vpm}. In particular, where the evasion applies, $A$ wins if and only if $\mathbf{x}_T\in\mathcal{D}_Q$ for some $Q\in\mathcal{A}^r$, and where the crossing on the surface applies, the $D$--$T$ team wins if and only if $\mathbf{x}_T\in\mathcal{E}_p$ for some $p\in\mathcal{C}^r$.
\end{corollary}

\begin{IEEEproof}
On $\Omega$, $V=V^-$ by \crefrange{thm:air}{thm:aircross}. Since $F_\zeta(Q)\ge z_Q$, only points with $z_Q<R$ matter, and with $\mathbf{q}_\zeta=(x_Q,y_Q,0)-\gamma d_A(\cos\zeta,\sin\zeta,0)$,
\begin{equation}\label{eq:diskzeta}
\begin{gathered}
F_\zeta(Q)<R\iff\|\mathbf{x}_T-\mathbf{q}_\zeta\|_2<\sqrt{R^2-z_Q^2}\\
\bigcap_{\zeta}\Bigl\{\mathbf{x}_T\in\mathcal{S}\Bigm|\|\mathbf{x}_T-\mathbf{q}_\zeta\|_2<\sqrt{R^2-z_Q^2}\Bigr\}=\mathcal{D}_Q,
\end{gathered}
\end{equation}
since the centers $\mathbf{q}_\zeta$ run over a circle of radius $\gamma d_A$ about the foot of $Q$. Thus $V^-<R$ if and only if, for every $\zeta$, $A$ meets $T$ or $\mathbf{x}_T$ lies in the first set of \eqref{eq:diskzeta} for some $Q$. Where the evasion applies, $V=\min_QJ(Q)$ with $J(Q)=\max_\zeta F_\zeta(Q)$, and $J(Q)<R$ holds if and only if $(d_T+\gamma d_A)^2+z_Q^2<R^2$, that is, $d_T<\sqrt{R^2-z_Q^2}-\gamma d_A$, or $\mathbf{x}_T\in\mathcal{D}_Q$ by \eqref{eq:diskzeta}. Where the crossing on the surface applies, $V=\max_pK(p)$, and $K(p)>R$ holds if and only if $d_T<\gamma d_A-R$. \Cref{thm:barrier} completes the proof.
\end{IEEEproof}

\begin{algorithm}[!t]
\caption{Barrier of the game of kind on the target plane}\label{alg:barrier}
\begin{algorithmic}[1]
\REQUIRE $\mathbf{x}_A$, $\mathbf{x}_D$, $\gamma$, $R$ and $r$
\STATE Sample the part of $\mathcal{A}^r$ with $z<R$ from \eqref{eq:rho}, and sample $\mathcal{C}^r$ from \eqref{eq:trace}.
\STATE For each sample $Q$, keep $\mathcal{D}_Q$ in \eqref{eq:disks} if $\varrho_Q>0$; for each $p$, keep $\mathcal{E}_p$ if $\varrho_p>0$.
\STATE Where the evasion applies, the winning region of $A$ is the union of the kept $\mathcal{D}_Q$; where the crossing on the surface applies, that of the $D$--$T$ team is the union of the kept $\mathcal{E}_p$; elsewhere, evaluate $V^-$ in \eqref{eq:Vpm} on a grid of $\mathbf{x}_T$.
\STATE Check \eqref{eq:monotone} along the boundary of these unions.
\ENSURE The barrier $\mathcal{B}$ on the target plane, as the boundary of the unions.
\end{algorithmic}
\end{algorithm}

\begin{figure}[!t]
\centering
\includegraphics[width=\columnwidth]{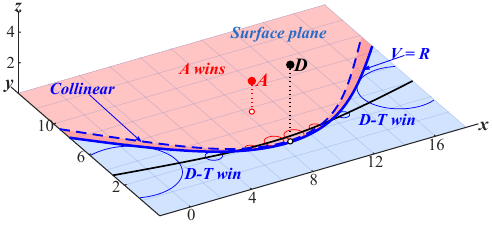}
\caption{Barrier on the target plane for $\mathbf{x}_A=(8.4,5.6,2)$, $\mathbf{x}_D=(8.6,1.6,5)$, $\gamma=0.6$, $r=1$ and $R=3$. The solid curve $V=R$ is the envelope of the capture disks of $A$ (red) and the escape disks of $T$ (blue) centered on the trace $\mathcal{C}^r$ (black); it separates the winning regions of $A$ (red) and of the $D$--$T$ team (blue). The dashed curve is the barrier of the collinear construction, which assigns the dark red strip to the $D$--$T$ team although $A$ wins there.}
\label{fig:barrier}
\end{figure}

\begin{remark}\label{rem:disks}
On $\mathcal{C}^r$, $z_Q=0$ and the two radii in \eqref{eq:disks} are $R-\gamma d_A$ and $\gamma d_A-R$: a point of the trace carries a capture disk if $A$ reaches it in less than $R/\gamma$, and an escape disk otherwise (\cref{fig:barrier}). \Cref{alg:barrier} only samples surfaces and curves; a Hamilton--Jacobi reachability computation~\cite{mitchell2005time,margellos2011hamilton} would have to cover the eight-dimensional state space. Outside $\Omega$, where \eqref{eq:monotone} fails, $V^+$ still bounds the value from above, and the disks $\mathcal{D}_Q$ give an inner estimate of the winning region of $A$.
\end{remark}

The barrier of~\cite{shi2025determination,shi2026constructing} is built differently. A point $p\in\mathcal{S}$ is safe for $T$ if $T$ reaches it before $A$ comes within $R$ of it, and if a straight strike of $A$ at $p$ is intercepted before $A$ comes within $R$ of $p$, that is, before the point $I$ at the distance $d_A-R$ from $\mathbf{x}_A$ on the segment to $p$. With $\rho(\varphi,\psi)$ in \eqref{eq:rho} taken along this segment, the two conditions read $g_1(p)>0$ and $g_2(p)>0$, where
\begin{equation}\label{eq:g12}
g_1(p)=d_A-R-\frac{d_T}{\gamma},\qquad g_2(p)=d_A-R-\rho,
\end{equation}
and the barrier of the collinear construction is the surface $B_c(\mathbf{x})=0$, with
\begin{equation}\label{eq:Bc}
B_c(\mathbf{x})=\max_{p\in\mathcal{S}}\,\min\bigl\{g_1(p),\,g_2(p)\bigr\}.
\end{equation}
In the plane, $\{g_1=0\}$ is the Apollonius oval of $T$--$A$, $\{g_2=0\}$ is the limit interception curve, and $B_c=0$ is their tangency~\cite[Thm.~4]{shi2026constructing}.

\begin{proposition}\label{prop:collinear}
Let $B_c(\mathbf{x})=0$ be attained at a nondegenerate $p^*$ with $g_1(p^*)=g_2(p^*)=0$ and $L=\|p^*-\mathbf{x}_A\|_2$. Let $(e_x,e_y,e_z)$ and $(\upsilon_x,\upsilon_y,\upsilon_z)$ be the direction cosines of $p^*-\mathbf{x}_A$ and of $I-\mathbf{x}_D$, $c=e_x\upsilon_x+e_y\upsilon_y+e_z\upsilon_z$, and let $\lambda\in[0,1]$ be the multiplier of $g_1$ at $p^*$. Then, with $k=(1-\lambda)/(1-c)$,
\begin{equation}\label{eq:Hc}
\medmuskip=1.3mu\relax \thickmuskip=2.5mu\relax
\min_{\beta,\theta_A}\max_{\delta,\theta_D,\zeta}\tfrac{\partial B_c}{\partial\mathbf{x}}\cdot\mathbf{f}
=1+kc-\sqrt{(1+kc)^2+k^2\tfrac{R^2}{L^2}(1-c^2)},
\end{equation}
which is negative unless $R=0$, $k=0$ or $c=\pm1$. For $R>0$, the surface $B_c=0$ is therefore not semipermeable.
\end{proposition}

\begin{IEEEproof}
At $p^*$, both constraints of the max--min are active, so the Karush--Kuhn--Tucker conditions give $\lambda\nabla_pg_1+(1-\lambda)\nabla_pg_2=0$, and the envelope theorem gives $\nabla B_c=\lambda\nabla g_1+(1-\lambda)\nabla g_2$. Let $\mathbf{e}=(e_x,e_y,e_z)$, $\bm{\upsilon}=(\upsilon_x,\upsilon_y,\upsilon_z)$ and
\begin{equation}\label{eq:hfun}
h(\rho,\mathbf{x})=\|\mathbf{x}_A+\rho\,\mathbf{e}-\mathbf{x}_D\|_2-\rho-r,
\end{equation}
which vanishes at $I=\mathbf{x}_A+\rho\,\mathbf{e}$. As $\partial\mathbf{e}/\partial\mathbf{x}_A=-(\mathbf{I}_3-\mathbf{e}\mathbf{e}^\top)/L$, its partial derivatives are
\begin{equation}\label{eq:hpartials}
\begin{aligned}
\frac{\partial h}{\partial\rho}&=c-1,\qquad\frac{\partial h}{\partial\mathbf{x}_D}=-\bm{\upsilon}\\
\frac{\partial h}{\partial\mathbf{x}_A}&=\bm{\upsilon}-\frac{\rho}{L}\bigl(\bm{\upsilon}-c\,\mathbf{e}\bigr),
\end{aligned}
\end{equation}
and the implicit function theorem~\cite{rudin1976principles} gives
\begin{equation}\label{eq:rhopartials}
\frac{\partial\rho}{\partial\mathbf{x}_D}=-\frac{\bm{\upsilon}}{1-c},\qquad
\frac{\partial\rho}{\partial\mathbf{x}_A}=\frac{\bm{\upsilon}-(\rho/L)(\bm{\upsilon}-c\,\mathbf{e})}{1-c}.
\end{equation}
With $\partial g_1/\partial\mathbf{x}_A=-\mathbf{e}$, $\partial g_2/\partial\mathbf{x}_A=-\mathbf{e}-\partial\rho/\partial\mathbf{x}_A$, $\partial g_2/\partial\mathbf{x}_D=-\partial\rho/\partial\mathbf{x}_D$ and $\rho=L-R$ at $p^*$, one obtains
\begin{equation}\label{eq:gradBc}
\begin{alignedat}{2}
\frac{\partial B_c}{\partial x_A}&=-(1+kc)e_x-k\frac{R}{L}(\upsilon_x-ce_x), &\quad \frac{\partial B_c}{\partial x_D}&=k\upsilon_x\\
\frac{\partial B_c}{\partial y_A}&=-(1+kc)e_y-k\frac{R}{L}(\upsilon_y-ce_y), &\quad \frac{\partial B_c}{\partial y_D}&=k\upsilon_y\\
\frac{\partial B_c}{\partial z_A}&=-(1+kc)e_z-k\frac{R}{L}(\upsilon_z-ce_z), &\quad \frac{\partial B_c}{\partial z_D}&=k\upsilon_z
\end{alignedat}
\end{equation}
and $\partial B_c/\partial x_T=(\lambda/\gamma)\hat{c}_x$, $\partial B_c/\partial y_T=(\lambda/\gamma)\hat{c}_y$, where $(\hat{c}_x,\hat{c}_y)$ are the horizontal direction cosines of $p^*-\mathbf{x}_T$. The vector $(\upsilon_x-ce_x,\upsilon_y-ce_y,\upsilon_z-ce_z)$ is orthogonal to $(e_x,e_y,e_z)$ and has the length $\sqrt{1-c^2}$, so that, by \eqref{eq:gradBc},
\begin{equation}\label{eq:normsBc}
\begin{aligned}
\Bigl\|\frac{\partial B_c}{\partial\mathbf{x}_A}\Bigr\|_2^2&=(1+kc)^2+k^2\frac{R^2}{L^2}\bigl(1-c^2\bigr)\\
\Bigl\|\frac{\partial B_c}{\partial\mathbf{x}_D}\Bigr\|_2&=k\sqrt{\upsilon_x^2+\upsilon_y^2+\upsilon_z^2}=k\\
\gamma\Bigl\|\frac{\partial B_c}{\partial\mathbf{x}_T}\Bigr\|_2&=\lambda\sqrt{\hat{c}_x^2+\hat{c}_y^2}=\lambda.
\end{aligned}
\end{equation}
Hence
\begin{equation}\label{eq:Hc1}
\begin{aligned}
&\min_{\beta,\theta_A}\,\max_{\delta,\theta_D,\zeta}\frac{\partial B_c}{\partial\mathbf{x}}\cdot\mathbf{f}=-\sqrt{(1+kc)^2+k^2\frac{R^2}{L^2}\bigl(1-c^2\bigr)}\\
&\qquad+k+\lambda,
\end{aligned}
\end{equation}
and $k+\lambda=1+kc$ by the definition of $k$, which gives \eqref{eq:Hc}. The square root exceeds $1+kc$ unless $kR\sqrt{1-c^2}=0$.
\end{IEEEproof}

\begin{remark}\label{rem:collinear}
\Cref{prop:collinear} explains the error of the straight strike. Since a capture only requires the distance $R$, $A$ may veer away from $D$, along $(\upsilon_x-ce_x,\upsilon_y-ce_y,\upsilon_z-ce_z)$, and still strike $T$; the collinear construction ignores this freedom and overestimates the winning region of the $D$--$T$ team. For $R=0$, \eqref{eq:Hc} vanishes and the construction is exact, in agreement with~\cite{garcia2018cooperative,garcia2019design}. The collinearity of $A$, $I$ and the capture point, taken from the two-on-one pursuit of~\cite{wasz2019two}, holds for a pursuer that has no one to evade; in the TAD game, $A$ also evades $D$.
\end{remark}

\begin{example}\label{ex:barrier}
Let $\mathbf{x}_A=(8.4,5.6,2)$, $\mathbf{x}_D=(8.6,1.6,5)$, $\gamma=0.6$, $r=1$ and $R=3$. \Cref{fig:barrier} shows the barrier \eqref{eq:barrier} built by \cref{alg:barrier}. Near the foot of $D$ it runs outside $\mathcal{C}^r$, where $A$ strikes from the air, and further out it follows the escape disks of $T$. Along it, the left-hand sides of \eqref{eq:monotone} are at most $-0.80$ (evasion) and at least $0.18$ (crossing), so \cref{lem:monotone} applies. In the window of \cref{fig:barrier}, the winning region of $A$ covers $62.7\%$ of the plane, whereas the collinear construction gives $59.1\%$. At the point $(2.960,4.233,0)$ of the collinear barrier, $\lambda=0.6771$, and \eqref{eq:Hc} equals $-0.0190$, as does its finite-difference evaluation.
\end{example}

\subsection{Winning Regions}\label{subsec:winning}

In the planar game, the winning region of $A$ on the target plane is unbounded~\cite{shi2025determination,shi2026constructing}. The surface target changes this.

\begin{proposition}\label{prop:bounded}
If $d_{AD}<r$, the set of target positions $\mathbf{x}_T\in\mathcal{S}$ for which $V(\mathbf{x})<R$ is bounded.
\end{proposition}

\begin{IEEEproof}
By the equality case of \eqref{eq:affine}, a horizontal plane $z=h$ cuts $\mathcal{A}^r$ in the conic
\begin{equation}\label{eq:section}
\ell_h^2(x,y)=r^2\bigl[(x-x_A)^2+(y-y_A)^2+(h-z_A)^2\bigr],
\end{equation}
where $\ell_h=\ell+(z_A-z_D)h$ with $\ell_h\ge0$. In the form \eqref{eq:conic}, \eqref{eq:section} keeps $\varepsilon_{11}$, $\varepsilon_{12}$ and $\varepsilon_{22}$ of \eqref{eq:quad}, while, with $\kappa_h=\kappa+(z_A-z_D)h$,
\begin{equation}\label{eq:quadh}
\begin{aligned}
\varepsilon_{13}(h)&=\kappa_h(x_A-x_D)+r^2x_A\\
\varepsilon_{23}(h)&=\kappa_h(y_A-y_D)+r^2y_A\\
\varepsilon_{33}(h)&=\kappa_h^2-r^2\bigl[x_A^2+y_A^2+(h-z_A)^2\bigr].
\end{aligned}
\end{equation}
Hence the discriminant of \eqref{eq:section} is \eqref{eq:Delta} for every $h$. As $d_{AD}<r$, every such section is an ellipse, and the sections with $0\le h\le R$ form a bounded set. By \cref{lem:trace}, $\mathcal{C}^r$ is an ellipse, and $\mathcal{R}_{AE}\cap\mathcal{S}$ is bounded. By \eqref{eq:Vpm}, $V(\mathbf{x})<R$ implies $V^-(\mathbf{x})<R$, so that, for the heading $\zeta=0$, either $A$ meets $T$ inside $\mathcal{R}_{AE}\cap\mathcal{S}$, or $F_0(Q)<R$ for some $Q\in\mathcal{A}^r$ with $z_Q<R$, that is, $\|\mathbf{x}_T+\gamma d_A(1,0,0)-(x_Q,y_Q,0)\|_2<R$. Both conditions confine $\mathbf{x}_T$ to a bounded set, which completes the proof.
\end{IEEEproof}

\begin{proposition}\label{prop:umbrella}
Let $D$ lie directly below $A$, that is, $x_D=x_A$, $y_D=y_A$ and $z_A-z_D>r$. If
\begin{equation}\label{eq:umbrella}
R<\frac{z_A+z_D+r}{2},
\end{equation}
then $\mathbf{x}\in\mathcal{R}_D$ for every $\mathbf{x}_T\in\mathcal{S}$.
\end{proposition}

\begin{IEEEproof}
Let $m=(z_{A_0}+z_{D_0})/2$. While $z_A<2m$, $D$ moves to the mirror image of $A$ in the plane $z=m$,
\begin{equation}\label{eq:mirror}
\begin{alignedat}{3}
x_D&=x_A, &\qquad y_D&=y_A, &\qquad z_D&=2m-z_A\\
\dot{x}_D&=\dot{x}_A, &\qquad \dot{y}_D&=\dot{y}_A, &\qquad \dot{z}_D&=-\dot{z}_A,
\end{alignedat}
\end{equation}
which is admissible because $D$ copies the horizontal velocity of $A$ and reverses its vertical one, so that it moves at unit speed and stays in $\mathcal{H}$; while $z_A\ge2m$, $D$ follows the foot of $A$ on $\mathcal{S}$, and \eqref{eq:mirror} holds again when $A$ descends to $z_A=2m$. (At constant speed, $D$ circles within an arbitrarily small distance of the foot, which perturbs the bound below by an arbitrarily small amount; the strict inequality \eqref{eq:umbrella} absorbs it.) Under \eqref{eq:mirror}, $\|\mathbf{x}_A-\mathbf{x}_D\|_2=2(z_A-m)$, so $A$ is intercepted as soon as $z_A\le m+r/2=(z_{A_0}+z_{D_0}+r)/2$. Before interception, every $\mathbf{x}_T\in\mathcal{S}$ satisfies
\begin{equation}\label{eq:umbrella1}
\|\mathbf{x}_A-\mathbf{x}_T\|_2\ge z_A>\frac{z_{A_0}+z_{D_0}+r}{2}>R,
\end{equation}
and $\mathcal{M}_1$ is never reached.
\end{IEEEproof}

\begin{figure}[!t]
\centering
\includegraphics[width=\columnwidth]{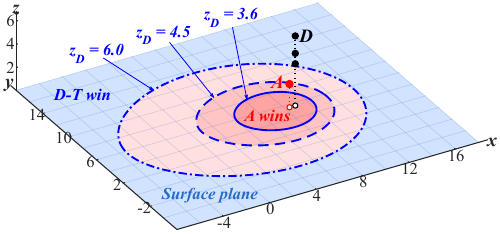}
\caption{Barriers on the target plane for $\mathbf{x}_A=(8.4,5.6,2)$, $\mathbf{x}_D=(8.9,5.6,z_D)$, $\gamma=0.6$, $r=1$ and $R=1.5$, with $D$ drawn at the heights $z_D=3.6$, $4.5$ and $6$ nearly above $A$. Since $d_{AD}=0.5<r$, each barrier is closed and the winning region of $A$ inside it (red, lighter for higher $D$) is bounded; it grows as $D$ climbs.}
\label{fig:bounded}
\end{figure}

\begin{remark}\label{rem:umbrella}
\Cref{prop:bounded,prop:umbrella} have no planar counterpart. A defender nearly above or below the attacker confines its strikes to a bounded part of the surface, and a defender directly below it, between the attacker and the surface, shields the whole surface. The bound \eqref{eq:umbrella} is the height of the lowest point of $\mathcal{A}^r$, so it cannot be improved. A defender above the attacker protects less, and the winning region of $A$ grows as $D$ climbs (\cref{fig:bounded}).
\end{remark}

\begin{example}\label{ex:bounded}
Let $\mathbf{x}_A=(8.4,5.6,2)$, $\gamma=0.6$, $r=1$ and $R=1.5$. For $\mathbf{x}_D=(8.9,5.6,z_D)$, $d_{AD}=0.5<r$ and, by \cref{prop:bounded}, the barrier is a closed curve (\cref{fig:bounded}). The winning region of $A$ has the area $27.5$, $76.0$ and $241.0$ for $z_D=3.6$, $4.5$ and $6$, and \eqref{eq:monotone} holds along the three barriers. For $\mathbf{x}_D=(8.4,5.6,0.5)$, directly below $A$, $(z_A+z_D+r)/2=1.75>R$, and $A$ cannot win from any target position.
\end{example}

\section{Conclusion}\label{sec:conclusion}

This article has studied a TAD game in which the target moves on a surface, while the attacker and the defender fly above it with non-zero capture radii. Firstly, the dominant regions of the players have been derived: the capture curve of $T$--$A$ is the trace of a Cartesian-oval surface, and the trace of the interception surface of $D$--$A$ is a conic whose type depends only on the horizontal distance between $A$ and $D$. Then, the game of degree has been solved by the two-sided maximum principle. The interception point lies in the air or on the surface, and the target either flees from its foot or runs under it. The strike directly above the target, a degenerate case in the plane, fails on an open region; there, and wherever the interception surface leans over the path of the target, the target runs under a point in the air. All solutions have been verified through the HJI equation and in closed loop. Finally, the barrier of the game of kind has been obtained as the level set $V=R$ of the value function. It corrects the collinear construction of earlier work for a non-zero capture radius, and it shows that a defender nearly above or below the attacker bounds or eliminates the winning region of the attacker, which has no planar counterpart.

Future work will treat the states outside the region of the straight-line solutions, where singular surfaces may arise, and will extend the analysis to unequal speeds of the attacker and the defender, to bounded turn rates, and to several attackers and defenders.

\bibliographystyle{IEEEtran-doi}
\bibliography{refs}

\begin{thebibliography}{10}
\providecommand{\url}[1]{#1}
\csname url@samestyle\endcsname
\providecommand{\newblock}{\relax}
\providecommand{\bibinfo}[2]{#2}
\providecommand{\BIBentrySTDinterwordspacing}{\spaceskip=0pt\relax}
\providecommand{\BIBentryALTinterwordstretchfactor}{4}
\providecommand{\BIBentryALTinterwordspacing}{\spaceskip=\fontdimen2\font plus
\BIBentryALTinterwordstretchfactor\fontdimen3\font minus \fontdimen4\font\relax}
\providecommand{\BIBforeignlanguage}[2]{{%
\expandafter\ifx\csname l@#1\endcsname\relax
\typeout{** WARNING: IEEEtran.bst: No hyphenation pattern has been}%
\typeout{** loaded for the language `#1'. Using the pattern for}%
\typeout{** the default language instead.}%
\else
\language=\csname l@#1\endcsname
\fi
#2}}
\providecommand{\BIBdecl}{\relax}
\BIBdecl

\bibitem{boyell1976defending}
R.~Boyell, ``Defending a moving target against missile or torpedo attack,'' \emph{IEEE Transactions on Aerospace and Electronic Systems}, vol. AES-12, no.~4, pp. 522--526, 1976, \doi{10.1109/taes.1976.308338}.

\bibitem{karasakal2011antiship}
O.~Karasakal, N.~E. {\"O}zdemirel, and L.~Kandiller, ``Anti-ship missile defense for a naval task group,'' \emph{Naval Research Logistics}, vol.~58, no.~3, pp. 304--321, 2011, \doi{10.1002/nav.20457}.

\bibitem{shaferman2010cooperative}
V.~Shaferman and T.~Shima, ``Cooperative multiple-model adaptive guidance for an aircraft defending missile,'' \emph{Journal of Guidance, Control, and Dynamics}, vol.~33, no.~6, pp. 1801--1813, 2010, \doi{10.2514/1.49515}.

\bibitem{shima2011optimal}
T.~Shima, ``Optimal cooperative pursuit and evasion strategies against a homing missile,'' \emph{Journal of Guidance, Control, and Dynamics}, vol.~34, no.~2, pp. 414--425, 2011, \doi{10.2514/1.51765}.

\bibitem{perelman2011cooperative}
A.~Perelman, T.~Shima, and I.~Rusnak, ``Cooperative differential games strategies for active aircraft protection from a homing missile,'' \emph{Journal of Guidance, Control, and Dynamics}, vol.~34, no.~3, pp. 761--773, 2011, \doi{10.2514/1.51611}.

\bibitem{prokopov2013linear}
O.~Prokopov and T.~Shima, ``Linear quadratic optimal cooperative strategies for active aircraft protection,'' \emph{Journal of Guidance, Control, and Dynamics}, vol.~36, no.~3, pp. 753--764, 2013, \doi{10.2514/1.58531}.

\bibitem{weiss2016minimum}
M.~Weiss, T.~Shima, D.~Castaneda, and I.~Rusnak, ``Minimum effort intercept and evasion guidance algorithms for active aircraft defense,'' \emph{Journal of Guidance, Control, and Dynamics}, vol.~39, no.~10, pp. 2297--2311, 2016, \doi{10.2514/1.G000558}.

\bibitem{kumar2017cooperative}
S.~R. Kumar and T.~Shima, ``Cooperative nonlinear guidance strategies for aircraft defense,'' \emph{Journal of Guidance, Control, and Dynamics}, vol.~40, no.~1, pp. 124--138, 2017, \doi{10.2514/1.G000659}.

\bibitem{isaacs1965differential}
R.~Isaacs, \emph{Differential games}.\hskip 1em plus 0.5em minus 0.4em\relax New York: John Wiley \& Sons, 1965.

\bibitem{rusnak2005lady}
I.~Rusnak, ``The lady, the bandits and the body guards -- a two team dynamic game,'' \emph{IFAC Proceedings Volumes}, vol.~38, no.~1, pp. 441--446, 2005, \doi{10.3182/20050703-6-cz-1902.00935}.

\bibitem{pachter2014active}
M.~Pachter, E.~Garcia, and D.~W. Casbeer, ``Active target defense differential game,'' in \emph{2014 52nd Annual Allerton Conference on Communication, Control, and Computing (Allerton)}, 2014, pp. 46--53, \doi{10.1109/allerton.2014.7028434}.

\bibitem{garcia2014cooperative}
E.~Garcia, D.~W. Casbeer, K.~Pham, and M.~Pachter, ``Cooperative aircraft defense from an attacking missile,'' in \emph{53rd IEEE Conference on Decision and Control}, 2014, pp. 2926--2931, \doi{10.1109/cdc.2014.7039839}.

\bibitem{garcia2015cooperative}
E.~Garcia, D.~W. Casbeer, and M.~Pachter, ``Cooperative strategies for optimal aircraft defense from an attacking missile,'' \emph{Journal of Guidance, Control, and Dynamics}, vol.~38, no.~8, pp. 1510--1520, 2015, \doi{10.2514/1.g001083}.

\bibitem{garcia2019design}
------, ``Design and analysis of state-feedback optimal strategies for the differential game of active defense,'' \emph{IEEE Transactions on Automatic Control}, vol.~64, no.~2, pp. 553--568, 2019, \doi{10.1109/tac.2018.2828088}.

\bibitem{pachter2019toward}
M.~Pachter, E.~Garcia, and D.~W. Casbeer, ``Toward a solution of the active target defense differential game,'' \emph{Dynamic Games and Applications}, vol.~9, no.~1, pp. 165--216, 2019, \doi{10.1007/s13235-018-0250-1}.

\bibitem{garcia2021complete}
E.~Garcia, D.~W. Casbeer, and M.~Pachter, ``The complete differential game of active target defense,'' \emph{Journal of Optimization Theory and Applications}, vol. 191, no. 2--3, pp. 675--699, 2021, \doi{10.1007/s10957-021-01816-z}.

\bibitem{garcia2017fast}
------, ``Active target defence differential game: fast defender case,'' \emph{IET Control Theory \& Applications}, vol.~11, no.~17, pp. 2985--2993, 2017, \doi{10.1049/iet-cta.2017.0302}.

\bibitem{garcia2021superior}
E.~Garcia, ``Cooperative target protection from a superior attacker,'' \emph{Automatica}, vol. 131, p. 109696, 2021, \doi{10.1016/j.automatica.2021.109696}.

\bibitem{liang2019differential}
L.~Liang, F.~Deng, Z.~Peng, X.~Li, and W.~Zha, ``A differential game for cooperative target defense,'' \emph{Automatica}, vol. 102, pp. 58--71, 2019, \doi{10.1016/j.automatica.2018.12.034}.

\bibitem{liang2021role}
L.~Liang, F.~Deng, M.~Lu, and J.~Chen, ``Analysis of role switch for cooperative target defense differential game,'' \emph{IEEE Transactions on Automatic Control}, vol.~66, no.~2, pp. 902--909, 2021, \doi{10.1109/tac.2020.2987701}.

\bibitem{liang2022reconnaissance}
L.~Liang, F.~Deng, J.~Wang, M.~Lu, and J.~Chen, ``A reconnaissance penetration game with territorial-constrained defender,'' \emph{IEEE Transactions on Automatic Control}, vol.~67, no.~11, pp. 6295--6302, 2022, \doi{10.1109/tac.2022.3183034}.

\bibitem{vonmoll2023circular}
A.~Von~Moll, M.~Pachter, D.~Shishika, and Z.~Fuchs, ``Circular target defense differential games,'' \emph{IEEE Transactions on Automatic Control}, vol.~68, no.~7, pp. 4065--4078, 2023, \doi{10.1109/TAC.2022.3203357}.

\bibitem{mammadov2024unified}
K.~Mammadov, C.-C. Lim, and P.~Shi, ``Unified optimality criteria for the {Target--Attacker--Defender} game,'' \emph{Journal of Control and Decision}, vol.~11, no.~4, pp. 572--589, 2024, \doi{10.1080/23307706.2023.2293964}.

\bibitem{li2026analytical}
Z.~Li, X.~Li, X.~Yang, X.~Yu, and J.~Qiu, ``An analytical approach for target defense differential games with speed-varying players,'' \emph{IEEE Transactions on Systems, Man, and Cybernetics: Systems}, vol.~56, no.~3, pp. 1850--1861, 2026, \doi{10.1109/TSMC.2025.3650198}.

\bibitem{wang2026multiagent}
J.~Wang, D.~W.~C. Ho, X.~Jin, F.~Li, and Y.~Tang, ``Multiagent {Target--Attacker--Defender} differential games with anomalous defenders under limited perception,'' \emph{IEEE Transactions on Automatic Control}, vol.~71, no.~5, pp. 2841--2856, 2026, \doi{10.1109/TAC.2025.3631483}.

\bibitem{fu2023justification}
H.~Fu and H.~H.-T. Liu, ``Justification of the geometric solution of a target defense game with faster defenders and a convex target area using the {HJI} equation,'' \emph{Automatica}, vol. 149, p. 110811, 2023, \doi{10.1016/j.automatica.2022.110811}.

\bibitem{dorothy2024apollonius}
M.~Dorothy, D.~Maity, D.~Shishika, and A.~Von~Moll, ``One {Apollonius} circle is enough for many pursuit-evasion games,'' \emph{Automatica}, vol. 163, p. 111587, 2024, \doi{10.1016/j.automatica.2024.111587}.

\bibitem{yan2021lifeline}
R.~Yan, Z.~Shi, and Y.~Zhong, ``Optimal strategies for the lifeline differential game with limited lifetime,'' \emph{International Journal of Control}, vol.~94, no.~8, pp. 2238--2251, 2021, \doi{10.1080/00207179.2019.1698770}.

\bibitem{wasz2019two}
P.~Wasz, M.~Pachter, and K.~Pham, ``Two-on-one pursuit with a non-zero capture radius,'' in \emph{2019 27th Mediterranean Conference on Control and Automation (MED)}, 2019, pp. 577--582, \doi{10.1109/med.2019.8798546}.

\bibitem{shi2024strategy}
Y.~Shi, C.~Wang, and Z.~Ding, ``Strategy analysis and design of {Target-Attacker-Defender} game with non-zero capture radius for attacker vehicle,'' \emph{IEEE Transactions on Intelligent Vehicles}, 2024, early access. \doi{10.1109/TIV.2024.3399822}.

\bibitem{shi2025determination}
Y.~Shi, C.~Wang, and D.~Liang, ``Determination of barrier surface in {Target-Attacker-Defender} game with capture radius for superior pursuer,'' \emph{Neurocomputing}, vol. 638, p. 130156, 2025, \doi{10.1016/j.neucom.2025.130156}.

\bibitem{shi2026constructing}
Y.~Shi, C.~Wang, S.~Li, and E.~Tian, ``Constructing barrier functions and optimal control for differential games in active defense,'' \emph{Automatica}, vol. 188, p. 112942, 2026, \doi{10.1016/j.automatica.2026.112942}.

\bibitem{miloh1982note}
T.~Miloh, ``A note on three-dimensional pursuit-evasion game with bounded curvature,'' \emph{IEEE Transactions on Automatic Control}, vol.~27, no.~3, pp. 739--741, 1982, \doi{10.1109/tac.1982.1102992}.

\bibitem{ardema1987three}
M.~D. Ardema and N.~Rajan, ``An approach to three-dimensional aircraft pursuit-evasion,'' in \emph{Pursuit-Evasion Differential Games}, Y.~Yavin and M.~Pachter, Eds.\hskip 1em plus 0.5em minus 0.4em\relax Oxford: Pergamon Press, 1987, pp. 97--110, \doi{10.1016/b978-0-08-034862-9.50013-x}.

\bibitem{segal1995barrier}
A.~Segal and T.~Miloh, ``Barrier strategies and capture criteria in a {3D} pursuit-evasion differential game,'' \emph{Optimal Control Applications and Methods}, vol.~16, no.~5, pp. 321--340, 1995, \doi{10.1002/j.1099-1514.1995.tb00024.x}.

\bibitem{segal1996inertial}
------, ``Inertial space optimal trajectories of aerial three-dimensional pursuit-evasion differential game,'' \emph{Journal of Guidance, Control, and Dynamics}, vol.~19, no.~3, pp. 612--620, 1996, \doi{10.2514/3.21665}.

\bibitem{yan2019construction}
R.~Yan, Z.~Shi, and Y.~Zhong, ``Construction of the barrier for reach-avoid differential games in three-dimensional space with four equal-speed players,'' in \emph{2019 IEEE 58th Conference on Decision and Control (CDC)}, 2019, pp. 4067--4072, \doi{10.1109/cdc40024.2019.9029495}.

\bibitem{garcia2020three}
E.~Garcia, D.~W. Casbeer, and M.~Pachter, ``Optimal strategies for a class of multi-player reach-avoid differential games in {3D} space,'' \emph{IEEE Robotics and Automation Letters}, vol.~5, no.~3, pp. 4257--4264, 2020, \doi{10.1109/lra.2020.2994023}.

\bibitem{yan2022matching}
R.~Yan, X.~Duan, Z.~Shi, Y.~Zhong, and F.~Bullo, ``Matching-based capture strategies for {3D} heterogeneous multiplayer reach-avoid differential games,'' \emph{Automatica}, vol. 140, p. 110207, 2022, \doi{10.1016/j.automatica.2022.110207}.

\bibitem{yan2022guarding}
R.~Yan, Z.~Shi, and Y.~Zhong, ``Guarding a subspace in high-dimensional space with two defenders and one attacker,'' \emph{IEEE Transactions on Cybernetics}, vol.~52, no.~5, pp. 3998--4011, 2022, \doi{10.1109/tcyb.2020.3015031}.

\bibitem{chen2024equilibrium}
N.~Chen, L.~Li, and W.~Mao, ``Equilibrium strategy of the pursuit-evasion game in three-dimensional space,'' \emph{IEEE/CAA Journal of Automatica Sinica}, vol.~11, no.~2, pp. 446--458, 2024, \doi{10.1109/jas.2023.123996}.

\bibitem{chen2025cooperative}
L.~Chen and Y.~Xia, ``Cooperative capture strategy for a group of slower {UAVs} to intercept one faster intruder in {3-D} space,'' \emph{IEEE Transactions on Aerospace and Electronic Systems}, vol.~61, no.~4, pp. 8757--8769, 2025, \doi{10.1109/TAES.2025.3545780}.

\bibitem{li2022different}
S.~Li, C.~Wang, and G.~Xie, ``Pursuit-evasion differential games of players with different speeds in spaces of different dimensions,'' arXiv:2202.13522, 2022, \doi{10.48550/arXiv.2202.13522}.

\bibitem{lee2020perimeter}
E.~S. Lee, D.~Shishika, and V.~Kumar, ``Perimeter-defense game between aerial defender and ground intruder,'' in \emph{Proc. 59th IEEE Conference on Decision and Control (CDC)}, 2020, pp. 1530--1536, \doi{10.1109/CDC42340.2020.9304213}.

\bibitem{plante1971pursuit}
D.~Plante and A.~Levis, ``Pursuit-evasion games with state constraints: The reconnaissance game,'' in \emph{1971 IEEE Conference on Decision and Control}, 1971, pp. 443--448, \doi{10.1109/cdc.1971.271033}.

\bibitem{pachter1987halfplane}
M.~Pachter, ``Simple-motion pursuit-evasion in the half plane,'' in \emph{Pursuit-Evasion Differential Games}, Y.~Yavin and M.~Pachter, Eds.\hskip 1em plus 0.5em minus 0.4em\relax Oxford: Pergamon Press, 1987, pp. 69--82, \doi{10.1016/b978-0-08-034862-9.50011-6}.

\bibitem{ardema1985combat}
M.~D. Ardema, M.~Heymann, and N.~Rajan, ``Combat games,'' \emph{Journal of Optimization Theory and Applications}, vol.~46, no.~4, pp. 391--398, 1985, \doi{10.1007/BF00939144}.

\bibitem{garcia2018cooperative}
E.~Garcia, D.~W. Casbeer, Z.~E. Fuchs, and M.~Pachter, ``Cooperative missile guidance for active defense of air vehicles,'' \emph{IEEE Transactions on Aerospace and Electronic Systems}, vol.~54, no.~2, pp. 706--721, 2018, \doi{10.1109/taes.2017.2764269}.

\bibitem{lockwood1961book}
E.~H. Lockwood, \emph{A Book of Curves}.\hskip 1em plus 0.5em minus 0.4em\relax Cambridge: Cambridge University Press, 1961, \doi{10.1017/cbo9780511569340}.

\bibitem{audin2003geometry}
M.~Audin, \emph{Geometry}, ser. Universitext.\hskip 1em plus 0.5em minus 0.4em\relax Berlin: Springer, 2003, \doi{10.1007/978-3-642-56127-6}.

\bibitem{basar1998dynamic}
T.~Ba{\c{s}}ar and G.~J. Olsder, \emph{Dynamic noncooperative game theory}, 2nd~ed., ser. Classics in Applied Mathematics.\hskip 1em plus 0.5em minus 0.4em\relax Philadelphia, PA: SIAM, 1998, \doi{10.1137/1.9781611971132}.

\bibitem{hartl1995survey}
R.~F. Hartl, S.~P. Sethi, and R.~G. Vickson, ``A survey of the maximum principles for optimal control problems with state constraints,'' \emph{SIAM Review}, vol.~37, no.~2, pp. 181--218, 1995, \doi{10.1137/1037043}.

\bibitem{rudin1976principles}
W.~Rudin, \emph{Principles of mathematical analysis}, 3rd~ed.\hskip 1em plus 0.5em minus 0.4em\relax New York: McGraw-Hill, 1976.

\bibitem{milgrom2002envelope}
P.~Milgrom and I.~Segal, ``Envelope theorems for arbitrary choice sets,'' \emph{Econometrica}, vol.~70, no.~2, pp. 583--601, 2002, \doi{10.1111/1468-0262.00296}.

\bibitem{mitchell2005time}
I.~Mitchell, A.~Bayen, and C.~Tomlin, ``A time-dependent {Hamilton--Jacobi} formulation of reachable sets for continuous dynamic games,'' \emph{IEEE Transactions on Automatic Control}, vol.~50, no.~7, pp. 947--957, 2005, \doi{10.1109/tac.2005.851439}.

\bibitem{margellos2011hamilton}
K.~Margellos and J.~Lygeros, ``{Hamilton--Jacobi} formulation for reach-avoid differential games,'' \emph{IEEE Transactions on Automatic Control}, vol.~56, no.~8, pp. 1849--1861, 2011, \doi{10.1109/tac.2011.2105730}.

\end{thebibliography}

\begingroup
\makeatletter
\def\@IEEEBIOskipN{12pt}
\makeatother

\begin{IEEEbiography}[{\includegraphics[width=1in,height=1.25in,keepaspectratio]{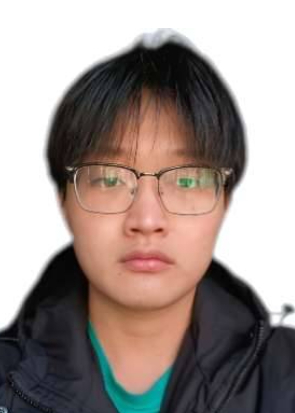}}]{Taotao Zhao}
(Student Member, IEEE) is pursuing the M.S. degree in systems science with the
Institute of Systems Science and the School of Automation, Qingdao University,
Qingdao, China. His research interests include differential games,
pursuit--evasion, active target defense, and Hamilton--Jacobi methods for the
design of feedback strategies in multiagent systems.
\end{IEEEbiography}

\begin{IEEEbiography}[{\includegraphics[width=1in,height=1.25in,keepaspectratio]{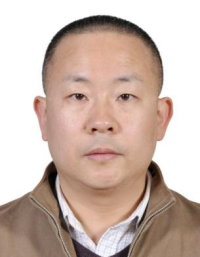}}]{Zhijian Ji}
(Senior Member, IEEE) received the M.S. degree in applied mathematics from
Ocean University of China, Qingdao, China, in 1998 and the Ph.D. degree in
control theory and control engineering from Peking University, Beijing, China,
in 2005. He is a Professor with the Institute of Systems Science and the School
of Automation, Qingdao University. His research interests include multiagent
systems, switched and hybrid systems, networked control, and distributed
coordination.
\end{IEEEbiography}

\begin{IEEEbiography}[{\includegraphics[width=1in,height=1.25in,keepaspectratio]{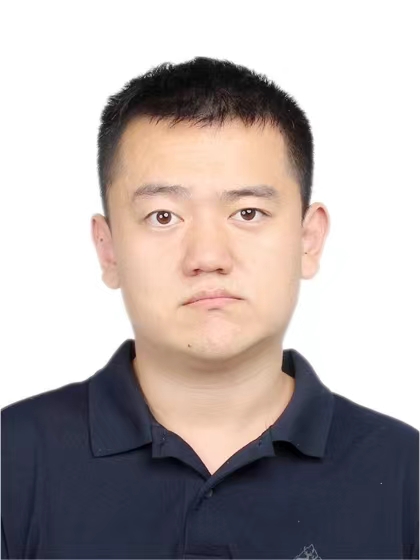}}]{Lanhao Zhao}
(Student Member, IEEE) received the B.E. degree in automation from Shandong
University of Science and Technology, China, in 2018 and the M.S. degree in
systems science from Qingdao University, China, in 2022. He is pursuing the
Ph.D. degree in control science and engineering at Beijing University of
Technology, Beijing, China. His research interests include multiagent systems,
complex networks, controllability and observability, and cooperative control.
\end{IEEEbiography}

\vspace*{\fill}
\endgroup

\end{document}